\documentclass[showpacs,onecolumn,aps,pra,longbibliography,superscriptaddress,notitlepage]{revtex4-2}
\usepackage{qcircuit}
\usepackage[dvips]{graphicx}
\usepackage{amsmath,amssymb,amsthm,mathrsfs,amsfonts,dsfont}
\usepackage{subfigure, epsfig}
\usepackage{braket}
\usepackage{svg}
\usepackage{bm}
\usepackage{enumerate}
\usepackage{wrapfig}
\usepackage{algpseudocode}
\usepackage{diagbox}
\usepackage{physics}
\usepackage{color}
\usepackage{multirow}
\usepackage[marginal]{footmisc}
\usepackage{comment}
\usepackage{makecell}
\usepackage[colorlinks = true]{hyperref}
\usepackage[nameinlink,capitalize]{cleveref}
\usepackage[ruled,vlined]{algorithm2e}

\usepackage{booktabs}

\newcommand{\ud}{\mathrm{d}}
\newcommand{\sgn}{\mathrm{sgn}}

\newcommand{\Or}{\mathcal{O}}

\newcommand{\mc}[1]{\mathcal{#1}}

\newtheorem{theorem}{Theorem}
\newtheorem{lemma}{Lemma}

\renewcommand{\sec}[1]{\hyperref[sec:#1]{Section~\ref*{sec:#1}}}
\newcommand{\app}[1]{\hyperref[app:#1]{Appendix~\ref*{app:#1}}}
\newcommand{\prob}[1]{\hyperref[prob:#1]{Problem~\ref*{prob:#1}}}
\newcommand{\thm}[1]{\hyperref[thm:#1]{Theorem~\ref*{thm:#1}}}
\newcommand{\lem}[1]{\hyperref[lem:#1]{Lemma~\ref*{lem:#1}}}
\newcommand{\cor}[1]{\hyperref[cor:#1]{Corollary~\ref*{cor:#1}}}
\newcommand{\alg}[1]{\hyperref[alg:#1]{Algorithm~\ref*{alg:#1}}}
\newcommand{\fgr}[1]{\hyperref[fgr:#1]{Figure~\ref*{fgr:#1}}}
\newcommand{\tab}[1]{\hyperref[tab:#1]{Table~\ref*{tab:#1}}}

\begin{document}
\title{
Circuit-Efficient Randomized Quantum Simulation of Non-Unitary Dynamics \\with Observable-Driven and Symmetry-Aware Designs
}

\begin{abstract}
We introduce \emph{random-LCHS}, a circuit-efficient randomized-compilation framework for simulating linear non-unitary dynamics of the form $\partial_t u(t) = -A(t)\,u(t) + b(t)$ built on the linear combination of Hamiltonian simulation (LCHS). We propose three related settings: the \emph{general random-LCHS} for time-dependent inhomogeneous linear dynamics; the \emph{observable-driven random-LCHS}, which targets estimation of an observable’s expectation at the final time; and the \emph{symmetric random-LCHS}, a time-independent, homogeneous reduction that can exploit physical symmetries. Our contributions are threefold: first, by randomizing the outer linear-combination-of-unitaries (LCU) layer as well as the deterministic inner Hamiltonian simulation layer, random-LCHS attains favorable resource overheads in the circuit design for early fault-tolerant devices; second, the observable-driven variant employs an unbiased Monte-Carlo estimator to target expectation values directly, reducing sample complexity; and third, integrating the physical symmetry in the model with the sampling scheme yields further empirical improvements, demonstrating tighter error bounds in realistic numerics.  We illustrate these techniques with theoretical guarantees as well as numerical verifications and discuss implementation trade-offs for near-term quantum hardware.
\end{abstract}

\author{Songqinghao Yang}
\email{sqhy2@cam.ac.uk}
\affiliation{Cavendish Lab., Department of Physics, University of Cambridge, Cambridge CB3 0HE, UK}

\author{Jin-Peng Liu}
\email{liujinpeng@tsinghua.edu.cn}
\affiliation{Yau Mathematical Sciences Center, Tsinghua University, Beijing 100084, China}
\affiliation{Yanqi Lake Beijing Institute of Mathematical Sciences and Applications, Beijing 100407, China}

\maketitle

\tableofcontents

\newpage

\section{Introduction}\label{sec:introduction}
Solving linear ordinary differential equations (ODEs) on a quantum computer is a foundational task with broad applications, from quantum dynamics~\cite{lloyd1996universal,berry2007efficient,childs2018toward} to general partial differential equations (PDEs)~\cite{childs2021high,an2025quantum}. In this work, we take as our starting point the initial value problem
$$
\frac{\mathrm{d}u(t)}{\mathrm{d}t} = -A(t)\,u(t) + b(t), \qquad u(0)=u_0,
$$
with $A(t)\in\mathbb{C}^{N\times N}$ and $u(t),b(t)\in\mathbb{C}^N$. Quantum algorithms aim to prepare an approximation to the normalized state $|u(T)\rangle = u(T)/\|u(T)\|$, from which global observables, overlaps, and other expectation values can be extracted efficiently. Extensions to nonlinear dynamics typically rely on linear-lift techniques, such as Carleman linearization~\cite{liu2021efficient, krovi2023improved, liu2023efficient, an2023quantumR}. The practical utility of quantum ODE solvers depends not only on asymptotic scalings in $T$, $\|A\|$, and the target precision $\epsilon$, but also on the chosen input model (oracle/block-encoding/Pauli decomposition), the cost of state preparation, and---crucially for early fault-tolerant hardware---the total qubit budget (computation + ancilla) as well as the complexity of required controlled operations~\cite{jennings2024cost,besedin2025realizing,google2025quantum}. In what follows, we focus on linear ODEs of the form above. Many asymptotically optimal constructions (e.g., post-Trotter simulation~\cite{low2017optimal,low2019hamiltonian}) obtain strong $T,\|A\|,\epsilon$ scaling at the price of large index registers, weight-loading circuits, and deep coherent control; on near-term and early fault-tolerant devices, these ancilla and control requirements can be the dominant practical bottleneck. Our design choice is to trade some of that coherent implementation and ancilla overhead for randomized sampling and simpler primitives: by replacing coherent layers with classical randomized selection, we reduce circuit width and control complexity at the expense of increased repetition (sampling) and a bias–variance resource tradeoff that is often preferable in realistic device regimes.

Randomized algorithms routinely replace costly global operations with numerous inexpensive local trials, whose results are combined statistically, a technique widely utilized in numerical linear algebra and optimization. In quantum simulation, sampling-based schemes such as qDrift~\cite{campbell2019random} demonstrate that stochastic compositions of small exponentials provide ancilla-free approximations to $e^{-iHt}$ with tight concentration bounds~\cite{chen2021concentration}. To improve simulation accuracy, qSWIFT~\cite{nakaji2024qswift} was proposed as a higher-order approach for qDrift; randomized-LCU schemes~\cite{chakraborty2024implementing,wang2024qubit} replace the explicit weight-loading process with random sampling. Building on these, Ref.~\cite{chakraborty2025quantum} worked on random-QSVT without block-encodings for which ancilla usage is reduced. Although it still exhibits suboptimal circuit depth and total gate complexity. These examples trade hardware-intensive resources for repetitions and classical post-processing---a tradeoff that is often favorable on near-term and early fault-tolerant platforms. 

Building on this perspective, we adopt the framework of linear combination of Hamiltonian simulation (LCHS)~\cite{an2023linear} for solving non-unitary dynamics integrated with a randomization strategy (random-LCHS). Unlike approaches based on dilation to linear systems or spectral mapping, LCHS achieves optimal state-preparation cost without such transformations. In our variant, both layers of the algorithm are randomized: the outer linear combination of unitaries (LCU) is implemented via Monte-Carlo sampling~\cite{chakraborty2024implementing}, and the inner Hamiltonian simulation subroutine is replaced with continuous qDrift (c-qDrift)~\cite{berry2020time}. This modification yields two key benefits. First, the ancilla overhead could collapse to nearly or exactly zero qubits, removing the need for large coherent weight registers and intricate controlled operations. Second, we evaluate the explicit gate complexity of the resulting circuits, rather than relying on abstract query complexity, which provides a more realistic measure of practical circuit-level costs. The tradeoff, as in other randomized methods, is statistical: estimator variance arises from both outer-node sampling and inner c-qDrift. Crucially, this variance is both quantifiable and reducible. By employing conserved–quantity–aware importance sampling (e.g., physDrift~\cite{yang2023randomized}) or unbiased node sampling~\cite{zhang2022unbiased}, we obtain provable concentration bounds that ensure favorable sample complexity and guarantee $\epsilon$-accurate results with high probability. These features make the random-LCHS framework particularly well-suited to align with the constraints and opportunities of early fault-tolerant quantum hardware. 

In the remainder of this section, we first review the historical development of quantum ODE solvers, then explain our motivation for combining the LCHS framework with randomized compilation techniques, and provide a detailed comparison with related work in the field.

\paragraph*{Development of quantum ODE solvers}
The literature on quantum ODE solvers naturally divides into families: linear-system-based methods (e.g.~\cite{berry2014high}), evolution-based methods (e.g.~\cite{fang2023time,an2025quantum}), and unitarization methods (e.g.~\cite{jin2024quantum,jin2023quantum,an2023linear}). Each family implicitly requires quantum access to structured $A(t)$ (e.g., sparse~\cite{berry2014high,berry2017quantum} or block-encoding~~\cite{an2025quantum}). The first family reduces the time-continuous problem to a high-dimensional linear system via time discretization (using multi-step methods, spectral methods, or truncated Taylor/Dyson expansions) and then applies quantum linear-system solvers (QLSAs). The canonical example is the multistep–HHL approach~\cite{berry2014high, harrow2009quantum}, later improved through higher-order discretizations and improved QLSAs~\cite{childs2017quantum,an2022quantum, lin2020optimal, subacsi2019quantum, costa2022optimal,jennings2023efficient,low2024quantum,low2024quantumlinear}. Assumptions for success in this family typically include efficient block-encoding of the dilated matrix, an acceptable condition number, and efficient preparation of the initial states. The complexity often scales with the condition number of the dilated system and depends on how the time discretization is performed. Subsequent works include truncated Taylor series~\cite{berry2017quantum,jennings2024cost}, truncated Dyson series~\cite{berry2024quantum,an2024fast}, spectral methods~\cite{childs2020quantum} and those used to solve linear partial differential equations~\cite{childs2021high}. The second family, named the evolution-based approach, constructs the propagator $e^{-At}$ (or $\mathcal{T}e^{-\int_0^T A(t)\,\mathrm{d}t}$) directly: time-marching methods sequentially encode the short-time integrators with amplitude-amplification~\cite{fang2023time}; QSVT-based methods encode the long-time integrators directly for time-independent parabolic and hyperbolic systems~\cite{an2025quantum}; and Lindbladian-based methods apply Lindbladian simulation for linear homogeneous ODEs~\cite{shang2024design}.

The third family of unitarization encompasses the Schr\"odingerisation~\cite {jin2023quantum,jin2024quantum} and linear combination of Hamiltonian simulation (LCHS)~\cite{an2023linear}. Among the reductions from non-unitary to unitary tasks, the LCHS is a particularly flexible and powerful technique. LCHS begins with the decomposition $A(t)=L(t)+iH(t)$, where $L(t)$ and $H(t)$ are Hermitian, and under mild stability conditions (for example $L(t)\succeq 0$) writes the time-ordered non-unitary propagator as a continuous linear combination of unitary propagators. A canonical representation is
$$
\mathcal{T}\exp\!\Big(-\int_0^T A(t)\,\mathrm{d}t\Big)
= \int_{\mathbb{R}} w(k)\,\mathcal{T}\exp\!\Big(-i\int_0^T \big(kL(t)+H(t)\big)\,\mathrm{d}t\Big)\,\mathrm{d}k,
$$
with an explicit weight function $w(k)$ \cite{an2023linear, an2025quantum}. Truncating and discretizing the integral yields a discrete LCU, $\sum_j w_j U_j$, with $U_j$ unitary evolutions under $k_jL(t)+H(t)$. The original LCHS method is closely related to Schr\"odingerisation~\cite{jin2023quantum,jin2024quantum}, a technique that converts non-unitary differential equations into a dilated Schr\"odinger equation by introducing an extra (momentum) dimension and imposing suitable initial conditions. The construction works in both qubit-based~\cite{jin2024quantumA, jin2024quantumB, jin2024quantumM, hu2025dilation} and continuous-variable settings~\cite{jin2024analog}. The naive LCHS can require costly coherent weight loading and force the simulation of Hamiltonians with large spectral norms if the integrand decays slowly. Recent families of LCHS identities remedy this by producing exponentially decaying integrands so that the truncation cutoff scales as $K=\mathcal{O}(\log(1/\epsilon))$, thereby achieving near-optimal matrix-query complexity \cite{an2023quantum, an2024laplace, novikau2025explicit,schleich2025arbitrary}. Further works build on this kernel-design approach to reduce truncation and circuit cost~\cite{low2025,an2025}. The LCHS theorem can also be extended to simulate time-evolution operators in infinite-dimensional spaces, including scenarios involving unbounded operators~\cite{lu2025infinite}.

While LCHS reduces non-unitary simulation to a continuum of Hamiltonian-simulation problems, the choice of inner Hamiltonian-simulation primitive profoundly affects practical resource costs. Traditional choices include Trotterization~\cite{lloyd1996universal,hatano2005finding,berry2007efficient,childs2021theory}, QSP/QSVT-based polynomial methods~\cite{low2017optimal,low2019hamiltonian}, interaction-picture simulation~\cite{low2018hamiltonian}, and randomized sampling schemes such as qDrift~\cite{campbell2019random} and its continuous analogue c-qDrift~\cite{berry2020time}. Randomized schemes approximate $e^{-iHt}$ by sampling terms from an explicit decomposition $H=\sum_\ell \alpha_\ell P_\ell$ according to weights and concatenating many small exponentials; qDrift gives unbiased estimators with concentration bounds, and c-qDrift extends this idea to parameterized families of Hamiltonians. Compared to QSVT-based approaches, randomized methods require almost no ancilla and avoid weight loading. Compared to Trotterization, they often need fewer distinct exponentials per logical timestep for large decompositions, and they do not scale explicitly with the system size.

\paragraph*{Random-LCHS} 
Our approach marries the LCHS reduction for non-unitary dynamics with randomized compilation at both algorithmic layers. Concretely, after the LCHS integral is truncated and discretized into quadrature nodes $\{k_j\}$ with weights $\{w_j\}$, there are two logical layers in which one may insert randomness: (i) the \emph{inner} Hamiltonian-simulation layer that implements each unitary block $U_j(T)=\mathcal{T}\exp\!\big(-i\int_0^T (k_j L(t)+H(t))\,\mathrm{d}t\big)$, and (ii) the \emph{outer} LCU/weight-loading layer that coherently prepares and combines the $U_j$ with amplitudes proportional to $\{w_j\}$. We pursue a \emph{hierarchy of randomization} that ranges from partial randomization (randomize the inner layer, Section~\ref{random_part1} Part 1) to complete randomization (randomize both inner and outer layers, Section~\ref{random_part2} Part 2). The high-level benefits are straightforward: randomness collapses coherent weight-loading and heavy ancilla requirements into sampling and classical post-processing, while the cost is statistical variance that can be quantified and driven down by importance sampling. 

State-preparation costs in random-LCHS are comparable to Trotter and qDrift baselines: both require preparing $\ket{u_0}$ and any inhomogeneity encodings $\ket{b(t)}$. Our framework preserves the same optimal state-preparation oracle queries $q := (\|u_0\|_1 + \|b\|_{L^1})/\|u(T)\|_1$ used in prior LCHS analyses~\cite{an2023quantum}; for homogeneous problems ($b\equiv 0$) this reduces to $q_0=\|u_0\|_1/\|u(T)\|_1$.

Now, we present the results for the three problems that will be the focus of this work. Throughout this work, we assume the normal settings (e.g., $L(t)\succeq 0$) for the LCHS framework without explicitly stating them. We formalize:

\vspace{0.2cm}
\medskip\noindent\textbf{Problem 1 (Linear Non-unitary Simulation).}  \textit{
Approximate $|u(T)\rangle$ for
$$
\frac{\mathrm{d}u(t)}{\mathrm{d}t} = -A(t)\,u(t) + b(t),\qquad u(0)=u_0.
$$
}
\vspace{0.2cm}

Throughout \textbf{Problem 1}, we generalize the input model considered in~\cite{berry2020time}, assuming the standard \emph{Block-encoding} (BE) access model~\cite{an2023quantum}. Many research efforts have tried to optimize block-encoding constructions for different classes of matrices, e.g., the general dense matrices~\cite{clader2023quantum, chakraborty2018power} and sparse matrices~\cite{camps2024explicit}. Here, it is worth emphasizing that for specific structured problems, the BE construction can sometimes be implemented in a surprisingly direct manner. In these cases, the apparent generality of BE does not translate into prohibitive complexity, and one can achieve relatively simple operator encodings without the full burden of the general LCU gadget~\cite{low2017optimal, gilyen2019quantum}. This also means that the input model here can generalize the sparse-matrix model considered in~\cite{berry2020time}, where $A(t)$ is a time-dependent $s$-sparse operator defined for $0\le t\le T$. We assume that for every $t$ the matrix $A(t)$ is $s$-sparse in the sense that the number of nonzero matrix elements in each row and each column is at most $s$, and that the locations of those nonzero entries do not change with time (only their values may depend on $t$). Under these assumptions, one may efficiently prepare the sparse-row and sparse-column superposition states.

\vspace{0.2cm}
\medskip\noindent\textbf{Result 1} [Informal Version of \textbf{Theorem~\ref{thm:complexity_inhomo_cqdrift} + Theorem~\ref{thm:ancilla-free-LCHS-inhomo}}]\label{result:complexity}\textit{
    The random-LCHS algorithm can produce the solution of \textbf{Problem 1}   $u(t)\in\mathbb{C}^{2^n}$ at time $T$ with error at most $\epsilon$ with state preparation cost $ \mathcal{O}\left(q\right)$ and an overall gate complexity $\widetilde{\mathcal{O}} \left( \frac{ n q^2 \norm{A}_{\infty,1}^2 }{\epsilon} \right)$ and $\log(\norm{L}_{\infty,\infty} T)$ ancilla. 
    Here $q=\frac{\|u_0\|_1+\|b\|_{L^1}}{\|u(T)\|_1}$, $\|A\|_{\infty,1}=\int_0^T\norm{A(\tau)}\mathrm{d}\tau$, $\|b\|_{L^1} = \int_0^T \|b(\tau)\|_1 \ud \tau$. If the input model is restricted to an $s$-sparse matrix, the gate complexity now becomes $\widetilde{\mathcal{O}} \left( \frac{ s^4 n q^2 \norm{A}_{\max,1}^2 }{\epsilon} \right)$. The algorithm can be modified as an ancilla-free variant with a higher statistical sampling cost.
    }
\vspace{0.2cm}

\noindent\textbf{Circuit efficiency:} The random-LCHS method preserves the optimal state-preparation cost of the original algorithm, requiring only a single copy of the input state per run. The dominant contribution to the sample-based Hamiltonian simulation subroutine of the operator $kL(t)+H(t)$, which we implement via the c-qDrift procedure. Alternatively, the coherent linear-combination-of-unitary design can be replaced with an ancilla-free variant (Theorem~\ref{thm:ancilla-free-LCHS-inhomo}). 
\begin{wrapfigure}{r}{0.5\textwidth}
  \begin{center}
    \includegraphics[width=0.48\textwidth]{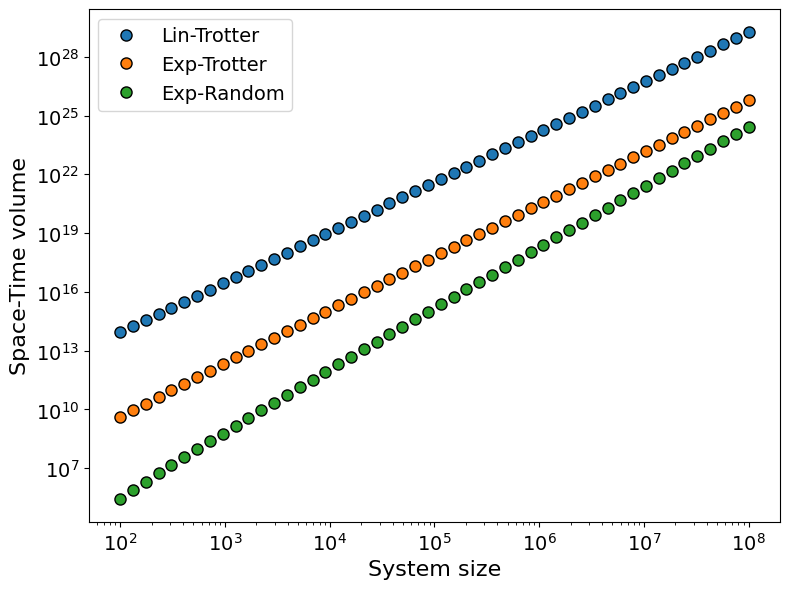}
  \end{center}
  \caption{Each curve (blue: the original linear-kernel + Trotterization approach; orange: Trotterization with the exponential kernel; green: our exponential-kernel approach) shows the space–time overhead as the simulated system size increases. The analytical curves are derived from simple asymptotic estimates of complexity and overhead, and they clearly illustrate the advantage of our method over Trotterization-based implementations. We do not compare our random-compiled method with approaches that use post-Trotter subroutines, since we restrict attention to near-term, practical fault-tolerant settings.
}
  \label{fgr:re}
\end{wrapfigure}
By avoiding the explicit LCU gadget and auxiliary registers, we trade hardware overhead for sampling variance: the algorithm requires more repetitions (Monte-Carlo-scaling $1/\epsilon^2$) to suppress statistical error. In this sense, the method is particularly well-suited to near-term hardware, where reducing circuit width is often more critical than minimizing variance. Moreover, the queries required in the Hamiltonian-simulation subroutine are structurally simpler than in deterministic Trotter and post-Trotter algorithms, making implementation easier in practice even though the asymptotic scaling is less favorable. Finally, although c-qDrift introduces a quadratic dependence on $\|A\|_{\infty,1}$, the $1$-norm dependence is believed to scale more gently with system size than the operator-norm dependence inherent to Trotterization, suggesting improved asymptotic performance in many-body settings. For example, if we consider the theoretical space-time volume (the product of wall-clock runtime and the number of physical qubits) per sample, for the Trotterized and randomized approaches, as in~\fgr{re}, the resulting randomized pipeline yields a reduction in space-time volume, thus improving the circuit efficiency.

We compare the complexities of random-LCHS and previous quantum algorithms for general time-dependent inhomogeneous linear ODEs in \tab{comparison_precise}. This table does not include some approaches that only handle time-independent ODEs~\cite{berry2017quantum,krovi2023improved,jennings2024cost,an2025quantum,low2024quantum,low2024quantumlinear}, homogeneous ODEs~\cite{shang2024design}, or dissipative ODEs~\cite{jennings2024cost,an2024fast,yang2025quantum}. Our method matches the near-optimal scaling in terms of the state preparation cost, and its query complexity depends on the tighter integrated spectral norm 
$\|A\|_{\infty,1}=\int_0^T \|A(t)\|\,\mathrm{d}t$ (an $L^1$-in-time quantity), which is strictly no larger than the worst-case bound 
$T\cdot\max_{t\in[0,T]}\|A(t)\|=T\|A\|_{\infty,\infty}$ employed in prior work. In our formulation, this bound lacks explicit time dependence. For certain systems, when large quantities of $\|A(t)\|$ only take a small amount of time interval, i.e.~$\int_0^T \|A(t)\| = o(T)$, the complexity of our algorithm can be \emph{independent} of $T$. 

\begin{table}[htb!]
\centering
\renewcommand{\arraystretch}{1.25}
\small
\begin{tabular}{l|c|c}
\toprule
\textbf{Method} & \textbf{Queries to Matrix Oracle} & \textbf{ Queries to State Preparation Oracles} \\ 
\midrule
\hline
Spectral method~\cite{childs2020quantum} &
$\widetilde{\Or}\!\big(qT\|A\|_{\infty,\infty}\,\mathrm{poly}\log(1/\epsilon)\big)$ &
$\mathcal{O}\!\big(qT\|A\|_{\infty,\infty} \,\mathrm{poly}\log(1/\epsilon)\big)$ \\ \hline

Truncated Dyson series~\cite{berry2024quantum} &
$\widetilde{\mathcal{O}}\!\big(qT\|A\|_{\infty,\infty} (\log(1/\epsilon))^2\big)$ &
$\mathcal{O}\!\big(qT\|A\|_{\infty,\infty} \log(1/\epsilon)\big)$ \\ \hline

Time-marching~\cite{fang2023time} &
$\widetilde{\Or}\!\big(qT^2\|A\|_{\infty,\infty}^2\log(1/\epsilon)\big)$ &
$\textcolor{red}{\mathcal{O}\!\big(q\big)}$ \\ \hline

Original LCHS~\cite{an2023linear} &
$\widetilde{\Or}\!\big(q^2T\|A\|_{\infty,\infty} /\epsilon\big)$ &
$\textcolor{red}{\mathcal{O}\!\big(q\big)}$ \\ \hline

Improved LCHS~\cite{an2023quantum,an2025} &
$\widetilde{\Or}\!\big(qT\|A\|_{\infty,\infty} (\log(1/\epsilon))^{1+o(1)}\big)$ &
$\textcolor{red}{\mathcal{O}\!\big(q\big)}$ \\ \hline

Optimal-scaling LCHS~\cite{low2025} &
$\widetilde{\Or}\!\big(qT\|A\|_{\infty,\infty} \log(1/\epsilon)\big)$ &
$\textcolor{red}{\mathcal{O}\!\big(q\big)}$ \\ \hline

\textbf{Random-LCHS (this work)} &
$\widetilde{\Or}\!\big(q^2\textcolor{blue}{\|A\|_{\infty,1}^2}/\epsilon\big)$ &
$\textcolor{red}{\mathcal{O}\!\big(q\big)}$ \\\hline
\bottomrule
\end{tabular}
\caption{
Asymptotic query complexities of matrix oracle and state-preparation oracles of quantum algorithms for time-dependent linear inhomogeneous ODEs. Here $T$ is the evolution time, $\epsilon$ is the error precision, and $q=\frac{\|u_0\|_1+\|b\|_{L^1}}{\|u(T)\|_1}$ is the rescaling factor of the solution.  $\|A\|_{\infty,1}=\int_0^T \|A(t)\|\,\mathrm{d}t$ and $\|A\|_{\infty,\infty}=\max_{t\in[0,T]}\|A(t)\|$ are norm estimates of $A(t)$, with the relation $\|A\|_{\infty,1} \le T\cdot\|A\|_{\infty,\infty}$. Assume the block-encoding normalization factor of $A(t)$ is $\norm{A(t)}$.}
\label{tab:comparison_precise}
\end{table}

In short, in \textbf{Problem 1}, when we leave the outer LCU (quadrature weights) intact but replace each deterministic inner simulation by c-qDrift, the practical improvements lie only in the norm scaling factor, compared to the Trotterization approach. The practical result is often a reduction in the circuit-depth-level efficiency per sample. When the algorithm admits a more hybrid implementation, i.e., randomizing the LCU, we can attain a zero ancilla regime, achieving both the per-sample circuit width and circuit depth efficiency. 

\vspace{0.2cm}
\medskip\noindent\textbf{Problem 2 (Observable-driven Simulation).}  \textit{
Given the linear ODE
$$
\frac{\mathrm{d}u(t)}{\mathrm{d}t} = -A(t)\,u(t) + b(t),\qquad u(0)=u_0,
$$
and an observable $O$, estimate the final-time expectation
$$
\mathbf{O} = u(T)^\dagger O\,u(T)
$$
to additive error $\epsilon$ with failure probability at most $\delta$.  
}
\vspace{0.2cm}

Here we assume two different input models: the Block-encoding model, and the Pauli model, which admits a Pauli decomposition $A(t)=\sum_l\alpha_l(t)A_l$, equivalent to the LCU model in~\cite{berry2020time}. The algorithm uses the LCHS quadrature to write $u(T)\approx\sum_{j}c_j U(T,k_j)u_0$ and then targets $\mathbf{O}$ via Monte-Carlo sampling of index pairs together with an inner randomized Hamiltonian-simulation subroutine. The algorithmic implementation is presented in \alg{obs_random_LCHS}.

\vspace{0.2cm}
\medskip\noindent\textbf{Result 2} [Informal Version of \textbf{Theorem~\ref{thm:ancilla_free_obs} + Theorem~\ref{unbiased_overhead}}]\label{thm:informal_observable_input_models}\textit{
Under the observable-driven random-LCHS framework for \textbf{Problem 2}, there exists a Monte-Carlo estimator that approximates $\mathbf{O}=u(T)^\dagger O u(T)$ to additive error $\epsilon$ with failure probability at most $\delta$. The practical circuit complexity scalings depend on the input model as follows.
}
\begin{enumerate}
\item \textit{\textbf{Block-encoding model:} Assume $A(t)$ is efficiently block-encoded. Using index-pair Monte-Carlo sampling and c-qDrift to implement each sampled evolution, one can achieve an additive error $\epsilon$ with failure probability $\le\delta$ using sample size $S$ and per-circuit segment count $r$ as
$$
S = \mathcal{O}\Big(\frac{\|O\|^{2}}{\epsilon^{2}}\log\!\frac{1}{\delta}\Big), \qquad r = \mathcal{O}\Big(\frac{nq^2\|A\|_{\infty,1}^{2}}{\epsilon}\Big).
$$
The total primitive cost is $O(S\cdot r)$. If $A(t)$ is $s$-sparse, we have $
r = \mathcal{O}\Big(\frac{q^2s^4n\|A\|_{\max,1}^{2}}{\epsilon}\Big)$.}
\item \textit{\textbf{Pauli input model:} Assume $A(t)=\sum_{l}\alpha_l(t)A_l$ is given in the Pauli/LCU form.  Using unbiased path sampling, sample size $S$ and per-circuit segment count $r$ scale as
$$
S = \mathcal{O}\big(1/\epsilon^{2}\big), \qquad r=\mathcal{O}\big(q^2\|\boldsymbol\alpha\|_{1,1}^{2}\big).
$$
Hence the overall sampled-segment cost is $\mathcal{O}\big(\|\boldsymbol\alpha\|_{1,1}^{2}\, q^2g_p/\epsilon^{2}\big)$, where $g_p$ is the gate cost per Pauli segment.
}
\end{enumerate}
\vspace{0.2cm}

The two approaches pursue fundamentally different information goals. Evolving the complete state $\ket{u(T)}$ is a reconstruction task: it produces a complete description from which any observable can subsequently be computed. Directly estimating a single expectation value $\mathbf{O}=u(T)^\dagger O u(T)$ is instead an inference task: it targets a specific functional of the evolution. It therefore permits tailored estimators that exploit the algebraic and statistical structure of that functional, often yielding simpler analysis and tighter guarantees for that particular quantity. The two approaches also differ in how errors propagate: a small operator-norm error in the prepared state controls all observables uniformly. In contrast, a small error in an estimator for $\mathbf{O}$ needs only to control that scalar; this means one can often trade uniform worst-case control for a more benign, task-oriented bias/variance allocation. Conceptually, observable tracing aligns with statistical estimation, whereas state evolution is a generative modeling problem whose success is measured by global distance metrics.

For \textbf{Problem~2}, the circuit efficiency arises not only from the randomized Hamiltonian-simulation subroutine using c-qDrift---which delivers the per-sample depth efficiency discussed in \textbf{Problem 1}---but also from our use of an unbiased sampler. The unbiased sampler further reduces the required circuit depth and makes the number of experimental repetitions independent of the target precision. Note that ancilla usage remains at most $\mathcal{O}(1)$.

\vspace{0.2cm}
\medskip\noindent\textbf{Problem 3 (Symmetry-preserving Simulation).} \textit{ 
Consider the homogeneous case and let $A\in\mathbb C^{2^n\times 2^n}$ be a time-independent generator admitting a local decomposition $A=\sum_r A_r$ where each $A_r$ is a Pauli string. The task is to prepare an $\epsilon$-approximation of the final state $\ket{u(T)}=e^{-AT}\ket{u_0}$ while preserving the model's conserved quantities and leveraging those symmetries to reduce effective sampling error.
}
\vspace{0.2cm}

We develop a symmetry-aware importance sampling scheme based on physDrift~\cite{yang2023randomized} together with random-LCHS for non-Hermitian system simulation with preserving physical symmetries. The approach can identify conserved quantities in the Hamiltonian models, and construct a symmetry-respecting sampling distribution for the randomized protocol. 

\vspace{0.2cm}
\medskip\noindent\textbf{Result 3} [Statement of \alg{outer-lchs-qdrift}]\label{thm:informal_symmetry}\textit{
Under the hypotheses of \textbf{Problem 3}, suppose the local decomposition $A=\sum_r a_rA_r$ is available and the model admits physical symmetries that pair local summands into symmetry-related partners (denote such a partner of $A_r$ by $A_r'$). Consider a symmetry-protected random-LCHS sampler formed by importance sampling from the LCHS kernel and by executing paired short evolutions of symmetry partners: sample pairs $(A_j, A_j')$ with weights proportional to $|a_j|+|a_j'|$.
}
\vspace{0.2cm}

The statement above formalizes a simple but broadly applicable design principle: enforce physical symmetries at the sampling rather than attempting post hoc corrections. Enforcing symmetry balance by pairing symmetry-related local summands changes only the micro-structure of each random macro-step (it preserves the marginal sampling probabilities) and therefore leaves big-$O$ circuit-level complexity unchanged; the practical benefit is a reduction of finite-sample error whenever symmetry-driven cancellations dominate the short-time dynamics. The algorithmic implementation is presented in \alg{outer-lchs-qdrift}. From an experimental and numerical standpoint, this manifests as a smaller error prefactor and often lowers the state error at fixed sampling budget.

We validated these assertions numerically on two benchmarking models: the complex Transverse-Field Ising Model (TFIM) and the Hatano-Nelson (HN) model. The results in~\fgr{expt} showcase average empirical improvements over the benchmark and~\fgr{ob} illustrate the empirical benefits with the symmetric implementation with observable-level evolution.

\paragraph*{Related-work}
Ref.~\cite{fang2025qubit} advances the practicality of quantum differential-equation solvers on early fault-tolerant hardware by introducing a Lindbladian-inspired, ancilla-sparing approach that embeds linear ODEs into an open-system construction and realizes their evolution using mid-circuit measurement, purification techniques, and other low-ancilla circuit patterns. The key feature of that proposal is its minimal ancilla footprint (often a single ancilla qubit), which makes it attractive for platforms where qubit count is the binding constraint. The manuscript reports a worst-case state-preparation oracle cost scaling as $\mathcal{O}(q^2)$, where $q$ denotes the natural LCHS state-preparation parameter $q=(\|u_0\|_1+\|b\|_{L^1})/\|u(T)\|_1$. Reaching the optimal $\mathcal{O}(q)$ query scaling would require introducing additional ancilla and amplitude-amplification style subroutines, a trade they avoid to keep the qubit budget minimal. We note that the general random-LCHS can achieve $\mathcal{O}(q)$ state preparation cost via amplitude-amplification; the ancilla-free random-LCHS also reports the $\mathcal{O}(q^2)$ state preparation scaling from Monte-Carlo sampling. In short, both the qubit-efficient approach and our ancilla-free random-LCHS try to minimize ancilla qubits at the cost of not optimal query dependence. 

Another work focuses on estimating expectation with LCHS using Monte-Carlo sampling~\cite{li2025dynamics}; this yields very low-ancilla, shallow circuits for direct observable estimation and is especially attractive when only a few observables are required and the problem has a structure that keeps sampling variance under control. By contrast, random-LCHS is designed to prepare the normalized final state directly, thereby supporting many downstream tasks. Practically, the two approaches trade the same resources in different ways---Ref.~\cite{li2025dynamics} trades potential variance for minimal ancilla when you only need limited information, while random-LCHS trades additional sampling for a usable quantum state.

For the rest of the paper: \sec{LCHS} introduces the Linear Combination of Hamiltonian Simulation formula and notations for norms; \sec{quantum_complexity}
proposes the random-LCHS algorithm for time-dependent inhomogeneous linear ODEs; \sec{observable-driven}
gives an observable-driven adaptation for estimating the expectations; \sec{symmetry_sampling} develops a symmetry-aware scheme for non-Hermitian systems with conserved quantities; \sec{discussion} concludes and discusses future directions.

\section{Preliminaries}\label{sec:LCHS}
\subsection{Linear Combination of Hamiltonian Simulation}
A key observation in the LCHS framework is that the propagator can be expressed using a family of kernel functions.  
For a kernel $f(k)$ we have
\begin{equation}\label{eqn:LCHS_improved}
    \mathcal{T} e^{-\int_0^T A(t)\, \ud t} 
    = \int_{\mathbb{R}} \frac{f(k)}{1-ik}\,
      \mathcal{T} e^{-i \int_0^T (kL(t)+H(t))\, \ud t}\, \ud k.
\end{equation}
This representation can be discretized and implemented via LCU.  
In the time-independent case, with $A \in \mathbb{C}^{2^{n}\times 2^{n}}$ and $\mathcal{R}(A)\succeq 0$, Theorem~5 of~\cite{an2023quantum} yields
\begin{equation}
    e^{-At} = \int_\mathbb{R} \frac{f(k)}{1-ik}\, e^{-i t(kL+H)} \ud k.
\end{equation}
This follows by setting $A(t)=A$ in Lemma~6 of~\cite{an2023quantum}, which reduces the Dyson series to $e^{-At}$ and ensures $\mathcal{P}\int_\mathbb{R} f(k) e^{-i t(kL+H)}\ud k = 0$.

In the original LCHS construction, $f(z)=1/(\pi(1+iz))$.  
Ref.~\cite{an2023quantum} proposed a faster-decaying kernel
\begin{equation}\label{eqn:kernel_exp}
    f(z) = \frac{1}{C_{\beta}\, e^{(1+iz)^{\beta}}}, \qquad 0<\beta<1,
\end{equation}
with normalization
\begin{equation}
C_\beta = 
\int_{\mathbb{R}} \frac{1}{(1-ik)e^{(1+ik)^{\beta}}} \ud k 
= i \int_{\mathbb{R}} \frac{e^{-(1+ik)^{\beta}}}{k+i}\, \ud k
= 2\pi e^{-2^\beta}.
\end{equation}
Here the $\beta$-power of $Re^{i\theta}$ ($\theta\in(-\pi,\pi]$) is defined by the principal value $R^\beta e^{i\beta\theta}$. 

Ideally, $f(k)$ should decay exponentially so that the truncation interval grows only logarithmically with the desired precision. The parameter $\beta$ makes this scaling slightly suboptimal, and~\cite{an2023quantum} shows that kernel design alone cannot remove this limitation. Eliminating the $\beta$-dependence of the truncation interval remains an open problem.

\subsection{Notation for norms}
To establish tight bounds on the random-LCHS framework, we first need to introduce the notation for norms. Here, we adopt the definition in the original c-qDrift script~\cite{berry2020time} based on the intuition: the complexity of simulating a time-dependent Hamiltonian should scale with its integrated size $ \int_0^T \|A(t)\|\,\mathrm{d}t $ (e.g., an $L^1$-type quantity) rather than the worst-case product $ T\max_t \|A(t)\| $. Our framework leverages the time structure of $A(t)$ to derive complexity bounds that depend on these integrated norms, thereby improving upon algorithms whose cost is driven solely by the uniform worst-case scenario.

For an $L$-vector $\boldsymbol{b}=(b_1,\dots,b_L)\in\mathbb{C}^L$ we use the standard $\ell_1$ and $\ell_\infty$ norms:
\begin{equation}\label{norm}
\norm{\boldsymbol{b}}_1=\sum_{j=1}^L|b_j|,\qquad
\norm{\boldsymbol{b}}_\infty=\max_{1\le j\le L}|b_j|.
\end{equation}
For a matrix $A$ we employ Schatten norms; in particular the trace (Schatten-1) norm and the spectral (Schatten-$\infty$) norm:
\begin{equation}
\norm{A}_1=\operatorname{Tr}\!\big(\sqrt{A^\dagger A}\big),\qquad
\norm{A}_2:=\sqrt{\mathrm{Tr}\big(A^\dagger A\big)},\qquad
\norm{A}=\norm{A}_\infty=\max_{\ket{\psi}}\|A\ket{\psi}\|_2.
\end{equation}
For a continuous scalar function $f:[0,t]\to\mathbb{C}$ we take the usual $L^1$ and $L^\infty$ norms on $[0,t]$:
\begin{equation}
\norm{f}_{L^1}=\int_0^T |f(t)|\,\mathrm{d}t,\qquad
\norm{f}_{L^\infty}=\max_{t\in[0,T]}|f(t)|.
\end{equation}
We explicitly consider two input models throughout this work: the block-encoding and sparse matrix model for the time-dependent system and the Pauli model for the time-independent system. 

\vspace{0.2cm}
\noindent\paragraph*{\textbf{(i) Block-encoding model and sparse matrix access:}} For the time-dependent operator $A(t)$, we use mixed norms $\|A\|_{p,q}$ defined by taking the Schatten-$p$ norm of the summand and the $L^q$ norm in time. The natural quantities controlling the simulation costs are the maximum norm:
\begin{equation}
\|A\|_{\max}:=\max_{j,k}|A_{j,k}|,\qquad
\|A\|_{\max,1}:=\int_0^T\|A(t)\|_{\max}\,\mathrm{d}t,\quad
\|A\|_{\max,\infty}=\max_{t}\|A(t)\|_{\max},
\end{equation}
with $\|A\|_{\max}\le\|A\|_\infty$. For example, in~\cite{berry2020time}, the gate complexity for c-qDrift for a $k$-local Hamitlonian with sparsity $s$ is $\widetilde{O}\big((s^2\norm{H}_{\max,1})^2k/\epsilon\big)$ and $\widetilde{O}\big((s^2\norm{H}_{\max,1})^2n/\epsilon\big)$ for a general $n$-qubit Hamiltonian. 

We do not assert a formal equivalence between the sparse-matrix access model and the block-encoding access model; they are distinct oracle models in general. For concreteness in our complexity statements, we adopt the common \emph{block-encoding} oracle as a convenient and widely used description of access to matrices (see e.g.~\cite{lin2022lecture}). The latter treats any matrix $A$ as its fundamental unit, provided we can efficiently implement a unitary that encodes $A$ in a specified subspace. Formally, a unitary $U_A$ is an $(\alpha, m, \epsilon)$-block-encoding of $A$ if:
\begin{equation}
\|A - \alpha (\bra{0^m} \otimes I) U_A (\ket{0^m} \otimes I) \| \leq \epsilon.
\end{equation}

When one is given sparse-oracle access instead of a block-encoding, standard constructions allow one to \emph{implement} a block-encoding of the sparse matrix (or otherwise simulate the sparse Hamiltonian) with additional algorithmic overhead.
We assume access to three oracles:
\begin{align*}
O_r\ket{\ell}\ket{i} = \ket{r(i,\ell)}\ket{i}, \qquad
O_c\ket{\ell}\ket{j} = \ket{c(j,\ell)}\ket{j}, \qquad
O_A\ket{0}\ket{i}\ket{j}=\left(A_{ij}\ket{0}+\sqrt{1-\abs{A_{ij}}^2}\right)\ket{i}\ket{j},
\end{align*}
where $r(i,\ell)$ and $c(j,\ell)$ give the column and row indices of the $\ell$-th nonzero element in row $i$ and column $j$, respectively. with the diffusion operator $D\ket{0^{n}}=\frac{1}{\sqrt{s}}\sum_{\ell\in[s]} \ket{\ell}$, one can easily show the encoding process for the sparse matrix $A$:
\begin{equation}
\bra{0}\bra{0^n}\bra{i}D^\dagger O_r^\dagger O_AO_cD\ket{0}\ket{0^n}\ket{j}=\frac{1}{\alpha}A_{ij}.
\end{equation}
We state complexity bounds in the block-encoding language for clarity, and remark where a sparse-oracle starting point changes prefactors or incurs conversion cost.

For a time-dependent matrix $A(t)$, one may construct a family of unitaries $U_A(t_k)$ by applying the block-encoding construction to a piecewise-constant approximation $A(t_k)$ at discrete times $t_k$. If the normalization is allowed to depend on time, we write $\alpha(t_k)$ and require that for each step.
$$
\big\| A(t_k) - \alpha(t_k)(\bra{0^m}\otimes I)U_A(t_k)\,(\ket{0^m}\otimes I)\big\| \le \epsilon_k .
$$
One can choose per-step tolerances $\epsilon_k$ so that the accumulated block-encoding error remains below the desired threshold.

Without loss of generality, we will assume $\alpha(t) = \norm{A(t)}$ throughout the paper. This would be helpful to simplify the complexity results relying on different norms (e.g.~$\|A\|_{\infty,1}$ and $\|A\|_{\infty,\infty}$), without referring to more redundant notations of $\alpha(t)$ (e.g.~$\norm{\alpha}_{1}$ and $\norm{\alpha}_{L^\infty}$).

\vspace{0.2cm}
\noindent\paragraph*{\textbf{(ii) Pauli model (homogeneous time-independent):}} The generator is a fixed linear combination of Pauli terms
\begin{equation}
A=\sum_{l=1}^L \alpha_l P_l, \qquad \|P_l\|_\infty \leq 1
\end{equation}
with constant, efficiently accessible coefficients $\alpha_l\in\mathbb{R}$. We use the standard coefficient norms from Eq.~\ref{norm}
\begin{equation}
\|\boldsymbol{\alpha}\|_1:=\sum_{l=1}^L|\alpha_l|,\qquad
\|\boldsymbol{\alpha}\|_\infty:=\max_{1\le l\le L}|\alpha_l|.
\end{equation}
These control the operator norm via
\begin{equation}\label{pauli}
\|A\|_\infty \le \sum_{l=1}^L|\alpha_{A_l}| = \|\boldsymbol{\alpha_A}\|_1,
\end{equation}
so that $\|A\|_\infty$ is bounded by the $\ell_1$ coefficient norm (while $\|\boldsymbol{\alpha}\|_\infty$ provides a per-term magnitude bound).

Finally, we write the general operator as
\begin{equation}
A(t)=L(t)+iH(t),
\end{equation}
and define pointwise and integrated spectral norms for each part:
\begin{equation}
\begin{split}
\|L\|_{\infty,1}:&=\int_0^T\|L(t)\|_{\infty}\,\mathrm{d}t,\quad
\|H\|_{\infty,1}:=\int_0^T\|H(t)\|_{\infty}\,\mathrm{d}t,\\
\|L\|_\infty \le& \sum_{j=1}^J|\alpha_{L_j}| = \|\boldsymbol{\alpha_L}\|_1,\quad
\|H\|_\infty \le \sum_{j=1}^J|\alpha_{H_j}| = \|\boldsymbol{\alpha_H}\|_1,
\end{split}
\end{equation}
Following~\cite{berry2020time}, to improve clarity, we express algorithmic costs directly in terms of the operator's actual norms (e.g., $ \|A\|_{\infty,1}$ and $\|\boldsymbol{\alpha_A}\|_{1}$) rather than looser upper bounds. We use standard asymptotic notation $\mathcal{O},\Omega,\Theta$ for asymptotic upper, lower, and tight bounds, and write $\widetilde{\mathcal{O}}(\cdot)$ to suppress polylogarithmic factors.

\section{General random-LCHS} \label{sec:quantum_complexity}
We now describe the quantum implementation of the random-LCHS scheme for a general time-dependent inhomogeneous linear ODE in the sparse-matrix model. For the solution of linear ODEs
\begin{equation}\label{eqn:ODE_solu}
    u(T) = \mathcal{T} e^{-\int_0^T A(t) \ud t} u_0 + \int_0^T \mathcal{T}e^{-\int_s^T A(t')\ud t'} b(t) \ud t,
\end{equation}
the quantum encoding of its discretized solution (derived in the Appendix) is approximated by
\begin{equation}\label{eqn:ODE_solu_discrete}
    u(T) \approx \sum_{j=0}^{M-1} c_j\, U(T,k_j)\ket{u_0}
    + \sum_{j'=0}^{M'-1}\sum_{j=0}^{M-1} c'_{j'} c_{j}\, U(T,t_{j'},k_{j})\ket{b(t_{j'})},
\end{equation}
where $U(T,t,k)=\mathcal{T}\exp\!\big(-i\int_t^T (kL(t')+H(t'))\,\mathrm{d}t'\big)$ and $U(T,k)$ denotes $U(T,0,k)$.  The first sum is the homogeneous evolution of the initial state $\ket{u_0}$; the double sum accounts for the inhomogeneous source terms $\ket{b(t_{j'})}$, with $c_j,c'_{j'}$ the discretization weights.

The quantum algorithm has two randomized layers. The inner layer is an efficient randomized simulator for the time-ordered unitary $U(T,s,k)$ (we implement this via c-qDrift). The outer layer replaces the usual LCU implementation by randomly sampling the discrete indices $(j,j')$ according to the weights $ c_j$ and $ c'_{j'}$.

\subsection*{Part 1}\label{random_part1}
We now present a randomized implementation of the LCHS method, using the continuous qDrift (c-qDrift) protocol as the Hamiltonian simulation subroutine. This approach offers a practical alternative to the deterministic method in the current literature~\cite{an2023quantum,an2023linear,an2024laplace,novikau2025explicit}, often resulting in a reduction in ancillary resource requirements as well as the circuit-depth-level efficiency.

For a given parameter $k_j$, we aim to simulate the time-dependent Hamiltonian $A^j(t) = H^j(t) + k_j L^j(t)$. We decompose it into its constituent terms: $H^j(t) = \sum_{\gamma=1}^{\Gamma} H^j_\gamma(t)$ and $L^j(t) = \sum_{\varrho=1}^{\rho} k_jL^j_\varrho(t)$, yielding $A^j(t) = \sum_{i=1}^{\Gamma + \rho} A^j_i(t)$. The c-qDrift protocol approximates the exact evolution channel $\mathcal{E}^j_T(\rho) = \mathcal{T}e^{-i\int_0^T A^j(t) \ud t} \rho \, \mathcal{T}e^{i\int_0^T A^j(t) \ud t}$ with a stochastic channel:
\begin{equation}\label{random_unitary}
\mathcal{U}_j(\rho)=\int_0^T\mathrm{d}t\, \sum_{i=1}^{\Gamma + \rho}p^j_i(t) \,
e^{-i \frac{A^j_i(t)}{p^j_i(t)}} \rho \,
e^{i \frac{A^j_i(t)}{p^j_i(t)}},
\end{equation}
where the time-dependent sampling probability for each term is defined as:
\begin{align}\label{dist_random}
p^j_i(t) &:=\frac{\norm{A^j_i(t)}_\infty}{\int_{0}^{T}\mathrm{d}s' \sum_{i=1}^{\Gamma+\rho} \norm{A^j_i(t')}_\infty} = \frac{\norm{A^j_i(t)}_\infty}{\sum_i\norm{A_i^j}_{\infty,1}}.
\end{align}
This distribution ensures that terms with larger norms are sampled more frequently. The error of this approximation is rigorously bounded by the following theorem, a proof of which can be found in~\cite{berry2020time}.

\begin{theorem}[Continuous qDrift Error Bound]\label{thm:cqDrift}
Let $A(t)=\sum_{l=1}^{L}A_l(t)$ be a time-dependent Hamiltonian defined for $0\leq t\leq T$. Assume each $A_l(t)$ is continuous and nonzero. Define the exact channel $\mathcal{E}_t(\rho) = E_t \rho E_t^\dagger$ where $E_t=\mathcal{T}e^{-i\int_{0}^{T} A(t) \ud t}$. Let $\mathcal{U}_t$ be the c-qDrift channel defined in Eq.~\eqref{random_unitary} with probabilities $p_l(t)=\norm{A_l(t)}_\infty / \sum_l\norm{A_l}_{\infty,1}$. Then,
	\begin{equation}
	\norm{\mathcal{E}_t-\mathcal{U}_t}_\diamond\leq 4\norm{A}_{\infty,1}^2 / r,
	\end{equation}
	where $r$ is the number of discrete time segments. To ensure a simulation error of at most $\epsilon$, it suffices to choose $r \geq \lceil 4 \norm{A}_{\infty,1}^2 / \epsilon \rceil$.
\end{theorem}
The linear combination of Hamiltonian simulations is then achieved by constructing a block-encoding of the sum $\sum_{j=0}^{M-1} c_j \, \mathcal{T}e^{-i\int_0^T A^j(t) \ud t}$. The standard method for this is the normal LCU technique. This coherent, ancilla-based approach provides a quantum circuit that implements the operator $\sum_{j=0}^{M-1} c_j U_j$ where $U_j = \mathcal{T}e^{-i\int_0^T A^j(t) \ud t}$, through the use of prepare and select oracles.

Here, we provide a brief introduction to the combination oracle. Interested reader can find more rigorous derivation in~\cite{an2023linear}: Let $O_c$ be a unitary that prepares the state $\sum_j \sqrt{c_j / \norm{\boldsymbol{c}}_1} \ket{j}$ on an ancilla register, and let $\text{SEL} = \sum_{j} \ket{j}\bra{j} \otimes U_j$ be the SELECT oracle that applies the unitary $U_j$ conditioned on the ancilla state $\ket{j}$. The standard LCU circuit, $(O_c^\dagger \otimes I) SEL (O_c \otimes I)$, then block-encodes the desired operator:
\begin{equation}
(\bra{0}_a \otimes I) \left[(O_c^\dagger \otimes I) SEL (O_c \otimes I)\right] (\ket{0}_a \otimes I) = \sum_{j=0}^{M-1}\frac{ c_j }{\norm{\boldsymbol{c}}_1}U_j.
\end{equation}
Measuring the ancilla register in the $\ket{0}$ state successfully applies the proportional operator to the system register. More concretely, for the coefficients $c_j$ in the quadrature formula, we assume access to a pair of state preparation oracles $(O_{c,l}, O_{c,r})$ acting as 
\begin{align*}
    O_{c,l}: \ket{0} \rightarrow \frac{1}{\sqrt{\|c\|_1}} \sum_{j=0}^{M-1} \overline{\sqrt{c_j}} \ket{j}, \qquad
    O_{c,r}: \ket{0} \rightarrow \frac{1}{\sqrt{\|c\|_1}} \sum_{j=0}^{M-1} \sqrt{c_j} \ket{j}. 
\end{align*}
Here for a complex number $z = Re^{i\theta}$, we have $\sqrt{z} = \sqrt{R} e^{i\theta/2}$ and its conjugate $\overline{z} = Re^{-i\theta}$. Furthermore, the access to the state preparation oracle acts as
\begin{equation}
    O_u: \ket{0} \rightarrow \ket{u_0}. 
\end{equation}
We now analyze the error for the homogeneous ODE ($b(t)=0$). The total error $\epsilon$ stems from two independent sources: The error $\epsilon_{\text{LCHS}}$ in the LCHS approximation of the true evolution by a linear combination of truncated discretization;
The error $\epsilon_{\text{cqd}}$ in implementing each individual channel $\mathcal{U}_j$ via c-qDrift. To bound the total error by $\epsilon$, we allocate $\epsilon/2$ to each source. The LCHS error is bounded by $\epsilon/2$ by choosing:
\begin{equation}
M = \mathcal{O}\left(\norm{A_L}_{\infty,\infty} T\log^{1+\frac{1}{\beta}}(\frac{1}{\epsilon})\right), \qquad \text{and}, \qquad K =\mathcal{O}\left(\log^{\frac{1}{\beta}}(\frac{1}{\epsilon})\right).
\end{equation}

The error for implementing a single c-qDrift channel $\mathcal{U}_j$ for time $T$ is governed by Theorem~\ref{thm:cqDrift}. For a channel approximating $\mathcal{T}e^{-i\int_0^T A^j(t) \ud t}$, the error is:
\begin{equation}\label{eqn:Trotter_error_td}
\norm{ \mathcal{U}_j - \mathcal{T}e^{-i\int_0^T A^j(t) \ud t} }_\diamond \leq \mathcal{O}\left( \frac{ (\norm{A^j}_{\infty,1})^2 }{r} \right).
\end{equation}
The error of the entire randomized LCHS approximation is then:
\begin{equation*}
\norm{ \sum_{j} c_j \mathcal{U}_j - \sum_{j} c_j \mathcal{T}e^{-i\int_0^T A^j(t) \ud t} } \leq \sum_{j} |c_j| \, \norm{ \mathcal{U}_j - \mathcal{T}e^{-i\int_0^T A^j(t) \ud t} } \leq \mathcal{O} \left( \norm{\boldsymbol{c}}_1 \frac{s^4\norm{A}_{\infty,1}^2}{r} \right).
\end{equation*}
To ensure this error is at most $\epsilon/2$, we require:
\begin{equation}\label{runs}
r = \mathcal{O} \left( \frac{ \norm{\boldsymbol{c}}_1 \norm{A}_{\infty,1}^2 }{\epsilon} \right).
\end{equation}

\begin{theorem}[Random-LCHS for Homogeneous ODEs]\label{thm:td_homo_random}
    Consider the homogeneous ODE with $b(t) \equiv 0$ and $A \in \mathbb{C}^{2^{n}\times 2^{n}}$ is s-sparse.
    Then, there exists a quantum algorithm that prepares an $\epsilon$-approximation of the state $\ket{u(T)}$, with gate complexity
    \begin{equation}
        \widetilde{\mathcal{O}} \left( \frac{ nq_0^2\norm{A}_{\infty,1}^2}{\epsilon} \right),
    \end{equation}
    where $q_0=\frac{\|u_0\|_1}{\|u(T)\|_1}$ and uses $\widetilde{\mathcal{O}}(\log(\norm{A_L}_{\infty,\infty} T))$ ancilla qubits.
\end{theorem}
\begin{proof}
    Let $V$ denote the $(\|\boldsymbol{c}\|_{1},\log(M),\epsilon')$-block-encoding of $\mc{T} e^{-\int_0^T A(t) \ud t}$. 
    We further write 
    \begin{equation}
        \|\boldsymbol{c}\|_1 \bra{0}_a V \ket{0}_a = \mc{T} e^{-\int_0^T A(t) \ud t} + E
    \end{equation}
    where $\|E\| \leq \epsilon'$. 
    We start with the state $\ket{0}_a\ket{0}$, where the ancilla register contains $\log(M)$ qubits. 
    After applying $O_{c}$ on the system register and $V$, we obtain the state 
    \begin{equation}
        V (I_a \otimes O_c) \ket{0}_a\ket{0} = \frac{1}{\|c\|_1} \ket{0}_a (\mc{T} e^{-\int_0^T A(t) \ud t})\ket{u_0} + \ket{\perp}. 
    \end{equation}

    Using the inequality $\|x/\|x\|-y/\|y\|\|\leq 2\|x-y\|/\|x\|$ for two vectors $x,y$, we can bound the error in the quantum state after a successful measurement as 
    \begin{equation}
        \left\|\ket{(\mc{T} e^{-\int_0^T A(t) \ud t}+E)u_0} - \ket{u(T)}\right\| \leq \frac{2\|Eu_0\|_1}{\|u(T)\|_1} \leq \frac{2\epsilon'\|u_0\|_1}{\|u(T)\|_1}. 
    \end{equation}
    To bound this error by $\epsilon$, it suffices to choose 
    \begin{equation}\label{aa}
        \epsilon' = \frac{\epsilon \|u(T)\|_1}{2\|u_0\|_1}. 
    \end{equation}
    Therefore, the complexity of a single run of our algorithm can be obtained by~\ref{runs} with $\epsilon'$, which becomes 
    \begin{equation}\label{runs_eps}
    \mathcal{O} \left(\left(\frac{\|u_0\|_1}{\|u(T)\|_1}\right) \frac{\|\boldsymbol{c}\|_{1}n\norm{A}_{\infty,1}^2}{\epsilon}  \right)=\mathcal{O} \left(\left(\frac{\|u_0\|_1}{\|u(T)\|_1}\right) \frac{n\norm{A}_{\infty,1}^2}{\epsilon}  \right).
    \end{equation}
    The expected number of repeats to get a success, after amplitude amplification, is
    \begin{equation}
        \mathcal{O}\left(\frac{\|\boldsymbol{c}\|_{1}\|u_0\|_1}{\|(\mc{T} e^{-\int_0^T A(t) \ud t}+E)u_0\|_1 }\right) \leq \mathcal{O}\left(\frac{\|\boldsymbol{c}\|_{1}\|u_0\|_1}{\|u(T)\|_1 - \epsilon' \|u_0\|_1 }\right) = \mathcal{O}\left(\frac{\|\boldsymbol{c}\|_{1}\|u_0\|_1}{\|u(T)\|_1} \right)=\mathcal{O}\left(\frac{\|u_0\|_1}{\|u(T)\|_1} \right). 
    \end{equation}
\end{proof}

The analysis for the general inhomogeneous ODE follows a similar path but requires handling the integration over the source term $b(t)$. The simulation requires approximating the operator $\int_0^T U(T, s, k) b(t) \ud t$, where $U(T, s, k)$ is the solution operator. This is done by discretizing the integral into $M'$ segments and applying the c-qDrift protocol as the subroutine for each segment $U(T, s_{j'}, k_j)$. For the LCU part, using the same approach as before, we may construct
\begin{equation}\label{eqn:algorithm_inhomo_sel_HS}
    \text{SEL}' = \sum_{j'=0}^{M'-1} \sum_{j=0}^{M-1} \ket{j'} \bra{j'} \otimes \ket{j} \bra{j} \otimes U(T,s_{j'},k_j) O_{b(s_{j'})}, \qquad \text{and}, \qquad  O_b = \sum_{j'=0}^{M'-1} \ket{j'}\bra{j'} \otimes O_{b(s_{j'})}.
\end{equation}
where the inhomogeneous oracle $O_b$ maps $\ket{0}$ to $\ket{b(s_{j'})}$. Applying the operator, $(O_{c',l}^{\dagger} \otimes O_{c,l}^{\dagger} \otimes I) \widetilde{\text{SEL}}' (O_{c',r} \otimes O_{c,r} \otimes I)$ to the zero state yields 
\begin{equation}\label{eqn:algorithm_general_inhomo_part}
    \frac{1}{\|c\|_1\|c'\|_1} \ket{0} \left( \sum_{j'=0}^{M'-1} \sum_{j=0}^{M-1} c_{j'}c_j U_{j,j'} \ket{b(s_{j'})} \right) + \ket{\perp}. 
\end{equation}

\begin{theorem}[Random-LCHS for Inhomogeneous ODEs]\label{thm:complexity_inhomo_cqdrift}
    Consider the inhomogeneous ODE system where $A \in \mathbb{C}^{2^{n}\times 2^{n}}$ is s-sparse.
    Suppose that $L(t)$ is positive semi-definite on $[0, T]$. 
    Let $q=\frac{\|u_0\|_1+\|b\|_{1}}{\|u(T)\|_1}$ and define $\Lambda = \sup_{p \geq 0, t \in [0,T]} \|A^{(p)}\|^{1/(p+1)} $ and $\Xi = \sup_{ p\geq 0, t \in [0,T] } \|b^{(p)}\|^{1/(p+1)} $, where the superscript $(p)$ indicates the $p$th-order time derivative. 
    Then we can prepare an $\epsilon$-approximation of the normalized solution $\ket{u(T)}$ with $\Omega(1)$ probability and a flag indicating success, by choosing 
    \begin{equation}
         M = \mathcal{O}\left( \norm{A_L}_{\infty,\infty} T\left(\log\left(\frac{q}{\epsilon}\right)\right)^{1+1/\beta} \right), \qquad M' = \widetilde{\mathcal{O}}\left( T (\Lambda+\Xi) \left(\log\left(\frac{q}{\epsilon}\right)\right)^{1+1/\beta} \right), 
    \end{equation}
    using $\mathcal{O}\left( q \right)$
        queries to the state preparation oracles, gate complexity
    \begin{equation}
            \widetilde{\mathcal{O}}( \frac{nq^2\norm{A}_{\infty,1}^2 }{\epsilon}),
        \end{equation}
       and
        $\widetilde{\mathcal{O}}\left(\log{\left( T\norm{L}_{\infty,\infty} (\Lambda+\Xi)\right)}\right)$
        extra ancilla qubits.
\end{theorem}
\begin{proof}
    Following Theorem~\ref{thm:td_homo_random} and Eq.~\ref{runs_eps}, we can prepare a single copy of the homogeneous state by querying $O_{c,l}, O_{c,r}, O_u$ $\mathcal{O}(1)$ times with gate count
    \begin{equation}
        \widetilde{\mathcal{O}}\left(\frac{  nq^2\norm{A}_{\infty,1}^2 }{\epsilon_1} \right).
    \end{equation}
    Similarly, we denote the gate complexity for preparation of the inhomogeneous state by querying $O_{c,l}, O_{c,r}, O_{c',l}, O_{c',r}, O_u, O_b$ $\mathcal{O}(1)$ as
    \begin{equation}
        \widetilde{\mathcal{O}}\left(\frac{  nq^2\norm{A}_{\infty,1}^2 }{\epsilon_2} \right).
    \end{equation}
    We now determine the choice of $\min\{\epsilon_1, \epsilon_2\}$. The error in the unnormalized solution can be bounded as 
    \begin{align*}
            &\left\| v - u(T)\right\|
             \leq \left\| \sum_{j} c_jU_j u_0 -   \mathcal{T}e^{-\int_0^T A(t) \ud t} u_0 \right\|+ \left\| \sum_{j,j'} c_{j'}c_j U_{j,j'} \ket{b(t_{j'})} -   \int_0^T \mathcal{T}e^{-\int_t^T A(t')\ud t'} b(t) \right\| \\
            & \leq \|c\|_1 \|u_0\|_1 \epsilon_1+ \|u_0\|_1 \left\| \sum_{j} c_j U(T,0,k_j) -   \mathcal{T}e^{-\int_0^T A(t) \ud t} \right\| \\
            & + \|c\|_1\|c'\|_1 \epsilon_2+ \left\|  \sum_{j,j'} c_{j'}c_j U(T,t_{j'},k_j) \ket{b(t_{j'})} -   \int_0^T \mathcal{T}e^{-\int_t^T A(t')\ud t'} b(t) \right\|. 
    \end{align*}
    Following~\cite{an2023linear}, we can bound the homogeneous (resp.~inhomogeneous) integral discretization error by $\epsilon_3/\|u_0\|$ (resp.~$\epsilon_3$) by choosing 
    \begin{equation}\label{eqn:proof_complexity_K}
        K = \mathcal{O}\left( \left(\log\left(\frac{\|u_0\|_1+\|b\|_{L^1}}{\epsilon_3}\right)\right)^{1/\beta} \right), \qquad M = \mathcal{O}\left( \norm{A_L}_{\infty,\infty}T\left(\log\left(\frac{\|u_0\|_1+\|b\|_{L^1}}{\epsilon_3}\right)\right)^{1+1/\beta} \right), 
    \end{equation}
    and 
    \begin{equation}
        M' = \widetilde{\mathcal{O}}\left( T (\Lambda+\Xi) \left(\log\left(\frac{1+\|b\|_{L^1}}{\epsilon_3}\right)\right)^{1+1/\beta}  \right). 
    \end{equation}
    Then we have 
    \begin{equation}\label{eqn:proof_complexity_inhomo_eq1}
        \left\| v - u(T)\right\| \leq \|c\|_1 \|u_0\|_1 \epsilon_1 +  \|c\|_1\|c'\|_1 \epsilon_2 + 2\epsilon_3, 
    \end{equation}
    and the error in the quantum state can be bounded as 
    \begin{equation}
        \left\| \ket{v} - \ket{u(T)}\right\| \leq \frac{2}{\|u(T)\|_1} \left\| v - u(T)\right\| \leq \frac{2\|c\|_1 \|u_0\|_1}{\|u(T)\|_1} \epsilon_1 +  \frac{2\|c\|_1\|c'\|_1}{\|u(T)\|_1} \epsilon_2 + \frac{4}{\|u(T)\|_1}\epsilon_3. 
    \end{equation}
    To bound this error by $\epsilon$, we can choose 
    \begin{equation}\label{eqn:proof_complexity_eps}
        \epsilon_1 = \frac{\|u(T)\|_1}{8\|c\|_1 \|u_0\|_1}\epsilon, \quad \epsilon_2 = \frac{\|u(T)\|_1}{8\|c\|_1\|c'\|_1}\epsilon, \quad \epsilon_3 = \frac{\|u(T)\|_1}{8}\epsilon,
    \end{equation}
    which dictates the number of runs we need for our sampling approach,
    \begin{align*}
        \widetilde{\mathcal{O}}&\left( \frac{  \norm{A}_{\infty,1}^2 }{\min \left\{\epsilon_1,\epsilon_2\right\}}\right) = \widetilde{\mathcal{O}}\left( \frac{  \norm{A}_{\infty,1}^2  \|c\|_1(\|u_0\|_1 + \|c'\|_1)}{ \|u(T)\|_1 \epsilon}\right) = \widetilde{\mathcal{O}}\left( \frac{  \norm{A}_{\infty,1}^2 (\|u_0\|_1 + \|c'\|_1)}{ \|u(T)\|_1 \epsilon}\right).
    \end{align*}
    The amplitude in the correct subspace is scaled by a factor of $\|v\|/(\|c\|_1(\|u_0\|_1+\|c'\|_1))$. 
    Consequently, using amplitude amplification, the number of algorithm repetitions required to achieve success is $\mathcal{O}(\|c\|_1(\|u_0\|_1+\|c'\|_1)/\|v\|)$. From the triangle inequality, we derive the lower bound:
    \begin{equation}
    \|v\| \geq \|u(T)\|_1 - \left(\|c\|_1 \|u_0\|_1 \epsilon_1 + \|c\|_1\|c'\|_1 \epsilon_2 + 2\epsilon_3\right) = \|u(T)\|_1 (1 - \epsilon).
    \end{equation}
    Substituting this into the complexity expression yields an extra prefactor:
    \begin{equation}\label{eqn:proof_complexity_repeat}
    \mathcal{O}\left( \frac{\|c\|_1(\|u_0\|_1+\|c'\|_1)}{\|v\|} \right) = \mathcal{O}\left( \frac{\|u_0\|_1 + \|b\|_{1}}{\|u(T)\|_1} \right).
    \end{equation}
    Finally, it is not hard to observe that the dominant factor for the ancilla comes from the weight-loading part, $\log(M)$ and $\log(M')$, which sums to the overall extra qubits.
\end{proof}

\subsection*{Part 2}\label{random_part2}
The standard Linear Combination of Unitaries provides a coherent method for implementing the operator sum $G=\sum_j c_j U_j$ but requires $\mathcal{O}(\log M)$ ancillary qubits to hold the control index. In the general LCHS framework, we demonstrated in the last section that more ancilla qubits are required for the extra inhomogeneous part, which adds further overhead for near-term devices. We present a randomized implementation that removes the outer layer of LCU at the cost of Monte-Carlo sampling overhead: indices are chosen classically according to a probability distribution derived from the coefficients, and the corresponding unitaries are executed on fresh shots. This reduces the quantum space complexity dramatically while introducing a sampling variance.

To preserve the complex linear combination within a probabilistic protocol, we sample indices $j$ from a distribution defined by the moduli of the coefficients, $p(j) = |c_j| / \norm{\boldsymbol{c}}_1$, where $\norm{\boldsymbol{c}}_1 = \sum_j |c_j|$. For each sampled index $j$, we apply the corresponding unitary $U_j$ (implemented via the c-qDrift protocol) multiplied by the factor $c_j / |c_j|$ to the initial state. This ensures the correct complex linear combination is achieved in expectation. The resulting channel for the homogeneous case is:
\begin{equation}
\mathcal{E}_{\text{rand}}^{\text{homo}}(\rho_0) = \sum_{j} p(j) \, \left(\frac{c_j}{|c_j|} U_j\right) \rho_0 \left(\frac{\overline{c_j}}{|c_j|} U_j^\dagger\right) = \frac{1}{\norm{\boldsymbol{c}}_1} \sum_{j} c_j U_j \rho_0 U_j^\dagger.
\end{equation}
Let $\bar{\rho} = \mathcal{E}_{\text{rand}}^{\text{homo}}(\rho_0)$ denote the average output state over many runs. From~\cite{chakraborty2024implementing}, we have the following bounds on the ancilla-free LCU.
\begin{theorem}[Randomized Unitary Sampling]{theorem}\label{thm:ancilla-free-LCU-theorem}
  Let $H\in \mathbb{C}^{N\times N}$ is a Hermitian matrix. Also let $\varepsilon\in (0,1)$ and suppose $F:[-1,1]\mapsto\mathbb{R}$ be some function such that
  $$
  \norm{F(H)-\sum_{j=1}^{M} c_j U_j}\leq \dfrac{\varepsilon}{3\norm{F(H)}},
  $$
 for some unitaries $U_j$ and $c_j\in \mathbb{R}\backslash\{0\}$ such that $\norm{\boldsymbol{c}}_1=\sum_{j}|c_j|\leq 1$. Suppose $V$ is a unitary sampled from the ensemble $\{c_j/\norm{\boldsymbol{c}}_1, U_j\}$, and applied to some initial state $\rho_0$. Then, the average density matrix, defined as
$$
\bar{\rho} = \mathbb{E}\left[V\rho_0V^{\dag}\right]=\dfrac{1}{\norm{\boldsymbol{c}}_1} \sum_{j=1}^{M} c_j U_j \rho_0 U^{\dag}_j,
$$
satisfies,
$$
\Tr\left[\Pi\bar{\rho}\right]\geq \Tr[\Pi F(H)\rho_0 F(H)^{\dag}] -\varepsilon,
$$
for any projector $\Pi$ acting on the state space of $\bar{\rho}$.
\end{theorem} 
It is not hard to see that, $F(A)$ in Theorem~\ref{thm:ancilla-free-LCU-theorem}, is just the time-order exponential integrator
\begin{equation}
    F(A) = \mathcal{T} e^{-\int_0^T A(t)\, \ud t} 
    = \int_{\mathbb{R}} \frac{f(k)}{1-ik}\,
      \mathcal{T} e^{-i \int_0^T (kL(t)+H(t))\, \ud t}\, \ud k.
\end{equation}
The function $F(A)$ is well defined since it is essentially depends on the Hermitian matrices $kL(t)+H(t)$. 

The single run complexity is presented in~\ref{runs_eps}. In regards to the Monte-Carlo importance sampling, the expected number of repeats is $\mathcal{O}\left(q_0^2/\epsilon^2 \right)$.

\begin{theorem}[Ancilla-Free LCHS for Homogeneous ODEs]
\label{thm:ancilla-free-LCHS-homo}
Consider the homogeneous ODE $\dot{u}(t) = -A(t)u(t)$ with $b(t) \equiv 0$ and $A \in \mathbb{C}^{2^{n}\times 2^{n}}$ is s-sparse. Let $G = \sum_{j=0}^{M-1} c_j U_j$ be an approximation to the ideal evolution operator $V(T) = \mathcal{T}e^{-\int_0^T A(t) \ud t}$, such that $\norm{G - V(T)} \leq \epsilon$. Let $\Pi$ be a projector onto the solution subspace. Then, for any initial pure state $\rho_0 = \ket{u_0}\bra{u_0}$, the average success probability of the randomized protocol satisfies:
\begin{equation}
\mathbb{E}[\Tr(\Pi \rho)] = \Tr[\Pi \bar{\rho}] \geq \frac{ |\braket{u(T)}{u_0}|^2 }{\norm{\boldsymbol{c}}_1^2} - \epsilon.
\end{equation}
There exists an ancilla-free randomized quantum algorithm that prepares an $\epsilon$-approximation of the state $\ket{u(T)}$ with $\Omega(1)$ success probability. The algorithm per-run requires one state-preparation-oracle query, a per-circuit gate complexity of
$\widetilde{\mathcal{O}} \left( \frac{ n  q_0^2\norm{A}_{\infty,1}^2 }{\epsilon} \right)$, and the Monte-Carlo sample size $\mathcal{O} \left( \frac{q_0^2}{\epsilon^2} \right)$, where $q_0 = \|u_0\|_1 / \|u(T)\|_1$. 
\end{theorem}

\begin{proof}
The lower bound on the expected success probability follows from the ancilla-free LCU Theorem~\cite{chakraborty2024implementing}, adapted for complex coefficients and the specific error bound $\epsilon$. The complexity is derived from the product of the number of c-qDrift segments $r$ required to achieve a simulation error $\epsilon_{\text{cqd}}$ (Eq.~\ref{runs}). The ancilla count stems from the fact that neither the classical sampling nor the c-qDrift protocol requires additional quantum circuit cost beyond those needed for the base oracles.
\end{proof}

Several complementary metrics are useful for assessing the performance of random-LCHS algorithms. First, the ancilla-qubit count---fewer ancillae improve feasibility on near-term devices. Second, the per-sample gate complexity, measured by the maximum circuit depth required for a single randomized Hamiltonian-simulation shot. Third, the total gate complexity, defined as the sum of the per-sample depths across all independent executions (i.e., the end-to-end cost after accounting for Monte-Carlo sampling). Together these metrics capture the key space–time tradeoffs introduced by replacing a coherent LCU with classical randomized sampling.

For the inhomogeneous case, $\dot{u}(t) = -A(t)u(t) + b(t)$, the solution involves simulating the operator $\int_0^T V(T, t) b(t) dt$. We approximate this via a double sum $\sum_{j'} \sum_{j} c_{j'} c_j U_{j,j'} \ket{b(t_{j'})}$. The ancilla-free protocol is extended by first sampling an index $j'$ for the time point from $p'(j') = |c_{j'}| / \norm{\boldsymbol{c'}}_1$, and then, conditioned on $j'$, sampling an index $j$ for the rescaling parameter from $p(j) = |c_j| / \norm{\boldsymbol{c}}_1$. 
\begin{theorem}[Ancilla-Free LCHS for Inhomogeneous ODEs]
\label{thm:ancilla-free-LCHS-inhomo}
Consider the inhomogeneous ODE system. Let $q = (\|u_0\|_1 + \|b\|_{L^1}) / \|u(T)\|_1$. Then, there exists a randomized quantum algorithm that prepares an $\epsilon$-approximation of the normalized solution $\ket{u(T)}$ with $\Omega(1)$ probability. The algorithm requires a per-circuit gate complexity of
$\widetilde{\mathcal{O}} \left( \frac{ n  q^2\norm{A}_{\infty,1}^2 }{\epsilon} \right)$, and the Monte-Carlo sample size $\mathcal{O} \left( \frac{q^2}{\epsilon^2} \right)$, where $q_0 = \|u_0\|_1 / \|u(T)\|_1$. 
\end{theorem}

The apparent two-stage sampling is equivalent to sampling a single composite index $k=(j,j')$ from the joint distribution $p(j,j')=\lvert c_j c_{j'}\rvert/(\lVert\boldsymbol{c}\rVert_1\lVert\boldsymbol{c'}\rVert_1)$. Writing $C=\lVert\boldsymbol{c}\rVert_1\lVert\boldsymbol{c'}\rVert_1=\|u_0\|_1+\|b\|_{L^1}$, the inhomogeneous contribution is the unnormalized vector $C=\sum_{j,j'} c_j c_{j'} U_{j,j'}\ket{b(s_{j'})}$, and a single sample produces the state proportional to $v/C$. Let $\ket{u(T)}$ be the full unnormalized solution and set $\gamma=\langle u(T)/\|u(T)\|\mid v/\|v\|\rangle$. The single-shot overlap with the desired normalized solution is $\alpha=\langle u(T)/\|u(T)\|\mid v/C\rangle=(\|u(T)\|/C)\gamma$, hence the naive postselection probability scales as $|\alpha|^2=(\|u(T)\|^2/C^2)|\gamma|^2$. The required number of sampler calls is $O(1/|\alpha|^2)$. Using the definition $q=(\|u_0\|_1+\|b\|_{L^1})/\|u(T)\|_1=C/\|u(T)\|_1$ shows that the normalization factor contributes quadratically in $q$, leading to an overall Monte-Carlo sampling number $\widetilde{\mathcal O}\big(q^2/\epsilon^2\big)$.

The ancilla-free approach achieves a radical reduction in space complexity, eliminating the $\mathcal{O}(\log M)$ ancilla qubit requirement inherent to the standard coherent LCU method. This advantage is particularly salient for early quantum devices where qubit count is a severely constrained resource. This benefit is achieved at the cost of repetitive circuit execution and classical post-processing. This tradeoff is characteristic of Monte-Carlo-type methods versus those employing coherent amplitude amplification.

\section{Observable-driven random-LCHS}\label{sec:observable-driven}
This section develops an observable-driven variant of the random-LCHS framework. Instead of preparing the full final state $\ket{u(T)}$, we adopt the same LCHS discretization and use the random compiler as a subroutine, focusing on estimating the expectation value
\begin{equation}
    \mathbf{O} := u(T)^\dagger O\, u(T),
\end{equation}
for a given observable (Hermitian operator) $O$. For near-term implementations, earlier work employed a hybrid approach based on importance sampling, utilizing the deterministic Hamiltonian simulation subroutine~\cite{an2023quantum,an2023linear}. 
We note that the improved $\mathcal{O}(\alpha_O/\epsilon)$, with $\alpha_O \geq \norm{O}$, scaling often cited for observable estimation relies on quantum amplitude estimation~\cite{brassard2000quantum}, which in turn requires coherent access to reflections about the target state to maintain the advantage~\cite{tang2025amplitude}. In the absence of such quantum operations, standard projective sampling exhibits the usual quadratic dependence on accuracy, $\mathcal{O}(\alpha_O^2/\epsilon^2)$. Our main results, therefore, present ancilla-free bounds and sample counts of order $S=\mathcal{O}\big(W^{2}\|O\|^{2}/\epsilon^{2}\big)$. In~\textit{Amplitude estimation} below, we briefly discuss how, under the stronger fault-tolerant assumption that one can implement the reflections, amplitude estimation reduces the sampling cost to the linear $\mathcal{O}(\alpha_O/\epsilon)$ scaling. In Part 2, our goal is to demonstrate improved observable estimation by replacing the c-qDrift-based random approach with unbiased sampling. We still consider the input model as a time-dependent block-encoded matrix $A(t)$. Our method employs two layers of sampling: outer Monte-Carlo sampling: A classical sampling procedure determines which randomized circuit is executed (same as~\cite{an2023quantum,an2023linear} with undeterministic subroutine); Inner unbiased gate-scaling decomposition: Each segment of the evolution is decomposed into randomly sampled gates with unbiased distribution.

\subsection*{Part 1}
We begin by recalling the LCHS representation for the final state (homogeneous case for simplicity):
\begin{equation}\label{eq:obs_lchs_decomp}
    u(T) \approx \sum_{j=0}^{M-1} c_j U(T,k_j)\ket{u_0},
\end{equation}
where $c_j\in\mathbb{C}$ are the quadrature weights from the LCHS discretization. Substituting Eq.~\ref{eq:obs_lchs_decomp} into the observable gives the double-sum representation used throughout this section:
\begin{equation}\label{eq:obs_double_sum}
    \mathbf{O} \approx \sum_{l=0}^{M-1}\sum_{j=0}^{M-1} \overline{c}_l c_j \; m_{l,j},\qquad m_{l,j}:=\braket{u_0|U(T,k_l)^\dagger \, O \, U(T,k_j)|u_0}.
\end{equation}
Because the coefficients $\overline{c}_l c_j$ are complex, we split the observable into its real and imaginary parts~\cite{an2023linear}:
\begin{align}\label{eq:obs_split}
    \mathcal{R}(\mathbf{O}) &= \sum_{l,j} |\Re(\overline{c}_l c_j)|\text{sgn}\big(\Re(\overline{c}_l c_j)\big)\; m_{l,j},\\
    \mathcal{I}(\mathbf{O}) &= \sum_{l,j} |\Im(\overline{c}_l c_j)|\text{sgn}\big(\Im(\overline{c}_l c_j)\big)\; m_{l,j}.
\end{align}
In the algorithmic constructions below we treat the two parts separately, sampling index pairs $(l,j)$ according to probability distributions proportional to the absolute weights $|\Re(\overline{c}_l c_j)|$ and $|\Im(\overline{c}_l c_j)|$, and combining the resulting unbiased estimates.

\begin{theorem}[Observable Estimation+C-qDrift]\label{thm:ancilla_free_obs}
Let $\mathbf{O}$ be defined in Eq.~\ref{eq:obs_double_sum}. Assume sampling access to approximate simulators $\widetilde U(T,k_j)$ via fresh shots, where each call, implemented with c-qDrift, has an error at most $\epsilon$. Assume further that measurements of $O$ on pure states yield outcomes bounded in magnitude by $\norm{O}$. 
Then there exists an ancilla-free Monte-Carlo estimator that returns an estimate $\widehat{\mathcal{O}}$ satisfying $\big|\widehat{\mathcal{O}} - \mathcal{O}\big| \le \epsilon$ with probability at least $1-\delta$, provided the number of independent index-pair samples $S$ and per-circuit segment count $r$ as
\begin{equation}\label{eq:ancilla_free_S_rewrite}
S = \mathcal{O}\Big(\frac{\|O\|^{2}}{\epsilon^{2}}\log\!\frac{1}{\delta}\Big), \qquad r = \mathcal{O}\Big(\frac{nq^2\|A\|_{\infty,1}^{2}}{\epsilon}\Big).
\end{equation}
\end{theorem}
We introduce $W_R,W_I$ in \alg{obs_random_LCHS} and denote
$$
W := \max\{W_R,W_I\} \le \|\boldsymbol{c}\|_1^2
$$
that collects the LCHS weight magnitude. The sample size has the general form $S = \mathcal{O}\Big(\frac{W^2\|O\|^{2}}{\epsilon^{2}}\log\!\frac{1}{\delta}\Big)$, in which $W^{2}$ can be ignored as $\mathcal{O}(\|\boldsymbol{c}\|_1^{4})=\mathcal{O}(1)$. The above theorem is easily proved with Hoeffding's bound,
\begin{equation}
\Pr\Big(|\overline{\mathbf{O}} - \mathbb{E}[\widehat{\mathbf{O}}]|>\epsilon/2\Big) \le 2\exp\Big(-\tfrac{2S\epsilon^2}{(2W\norm{O})^2}\Big).
\end{equation}
\begin{algorithm}[H]
\caption{Observable-driven random-LCHS with c-qDrift}
\label{alg:obs_random_LCHS}
\KwIn{Coefficients $\{c_j\}_{j=0}^{M-1}$, approximate simulators $\{\widetilde U(T,k_j)\}$, state-preparation $O_u$ for $\ket{u_0}$, measurement access to observable $O$, target error $\epsilon$, failure probability $\delta$, sample count $S$.}
\KwOut{Estimator $\widehat{\mathcal{O}}$ of the observable expectation.}
\BlankLine
\textbf{Preprocessing:}\;
Define weights $W^{(R)}_{l,j}:=|\Re(\overline{c}_l c_j)|$, \quad
$W^{(I)}_{l,j}:=|\Im(\overline{c}_l c_j)|$\;
Let $W_R=\sum_{l,j}W^{(R)}_{l,j}$ and $W_I=\sum_{l,j}W^{(I)}_{l,j}$\;
\BlankLine
\textbf{Main procedure:}\;
\For{$\mathrm{part}\in\{\mathrm{Re},\mathrm{Im}\}$}{
  Sample $S$ index pairs $(l_t,j_t)$ i.i.d.\ from
  $p_{l,j}^{(\mathrm{part})}=W^{(\mathrm{part})}_{l,j}/W_{\mathrm{part}}$\;
  \For{$t=1$ \KwTo $S$}{
    Prepare $\ket{u_0}$ and apply $\widetilde U(T,k_{j_t})$\;
    Measure via the c-qDrift subroutine to obtain estimator $\widehat m_{l_t,j_t}$ of $m_{l_t,j_t}$\;
    Form signed sample
    \[
      X_t^{(\mathrm{part})}
      = W_{\mathrm{part}}\cdot
        \sgn\!\big(\,(\mathrm{part})(\overline{c}_{l_t}c_{j_t})\big)\cdot
        \widehat m_{l_t,j_t},
    \]
    where for $\mathrm{part}=\mathrm{Re}$ we use $\sgn(\Re(\cdot))$ and for $\mathrm{part}=\mathrm{Im}$ we use $\sgn(\Im(\cdot))$\;
  }
  Compute empirical mean $\overline X^{(\mathrm{part})}=\frac{1}{S}\sum_{t=1}^S X_t^{(\mathrm{part})}$\;
}
\Return $\widehat{\mathbf{O}}=\overline X^{(R)} + i\,\overline X^{(I)}$\;
\end{algorithm}

Let $\delta'$ and $\epsilon'$ denote the upper bounds on the failure probability and on the c-qDrift evolution error, respectively. For long-time evolution with $r$ segments, the c-qDrift evolution error obeys the scaling
\begin{equation}\label{eq:cqdrift_error_repeat}
\big\|\rho^{\mathrm{qd}}-\rho^{\mathrm{ideal}}\big\| = O\!\left(\frac{n \|A\|_{\infty,1}^2}{r}\right),
\end{equation}
Hence, to ensure the c-qDrift contribution to the simulation error is at most $\epsilon'$, we must choose $r$ so that
\begin{equation}\label{eq:r_required}
\frac{n \|A\|_{\infty,1}^2}{r} \lesssim \epsilon'
\quad\Longrightarrow\quad
r = O\!\left(\frac{n  \|A\|_{\infty,1}^2}{\epsilon'}\right).
\end{equation}
Following~\cite{an2023quantum}, the bias induced by the c-qDrift approximation satisfies
\begin{equation}
\left|\|\boldsymbol{c}\|_1^2\overline{\mathbf{O}} - u(t)^{*} \mathbf{O} u(t)\right|
\;=\;\|c\|_1^2 \sum_{k,k'}\frac{W_{k,k'}}{\|c\|_1^2}\epsilon' \;=\; \|c\|_1^2 \epsilon'.
\end{equation}
To make this bias at most $\epsilon/2$, it suffices to set
\begin{equation}\label{eq:epsprime_choice}
\epsilon' = \frac{\epsilon}{2\|c\|_1^2}.
\end{equation}
Substituting Eq.~\ref{eq:epsprime_choice} into Eq.~\ref{eq:r_required}, with the state-prepartion amplitude factor in Eq.~\ref{aa}, gives the required number of c-qDrift segments
\begin{equation}\label{eq:r_final}
r = \mathcal{O}\left(\frac{nq_0^2  \|A\|_{\infty,1}^2}{\epsilon'}\right)
= \mathcal{O}\left(\frac{nq_0^2  \|A\|_{\infty,1}^2 \, \|c\|_1^2}{\epsilon}\right).
\end{equation}
For the success probability, the overall success probability after $S$ independent trials is at least $(1-\delta')^{S}\ge 1-S\delta'$. To guarantee an overall success probability $\ge 1-\delta/2$ we may choose $\delta' = \frac{\delta}{2S}$.

\vspace{0.2cm}
\textbf{Amplitude estimation.} In the fault-tolerant quantum computation scenario, we can apply the unbiased amplitude estimation with improved accuracy~\cite{brassard2000quantum,cornelissen2023sublinear,rall2023amplitude,van2023quantum}. The details refer to Theorem 12 of~\cite{brassard2000quantum} and Theorem 2.4 of~\cite{van2023quantum}.

Given the reflection operators $R_u = 2\ket{u(T)}\bra{u(T)} - I$ and $R_O = 2O - I$, where $O$ is the block-encoded observable, the quantum amplitude estimation algorithm can estimate $\bra{u(T)}O\ket{u(T)}$ within error $\epsilon$ with probability at least $1-\delta$, with the quantum sampling complexity
\begin{equation}\label{eq:amplitude_estimation}
    \sqrt{S} = \mathcal{O}\!\left( \frac{ \norm{O}}{\epsilon} \log\!\frac{1}{\delta} \right),
\end{equation}
i.e., the number of queries to the reflections $R_u$ and $R_O$. 

Here the operator $R_u$ indicates the capability of backtracking from $\ket{u(T)}$ to $\ket{u_0}$. It is generally challenging to invert the non-unitary and random processes, but it may be possible if we have a stronger protocol to invert the quantum channel, such as Block Encoding of Linear Transformation (BELT)~\cite{wei2025belt}.

\subsection*{Part 2}
We briefly introduce the unbiased random circuit compiler (URCC) sampling framework based on the Dyson series and Poisson-weighted path sampling~\cite{zhang2022unbiased}. This provides a general method for producing unbiased Monte-Carlo estimators for time-ordered evolution. However, here we will restrict ourselves to a different input model---the time-dependent Pauli model (the same as the LCU model in~\cite{berry2020time}, where the input matrix $A(t)$ admits decomposition $A(t)=\sum_{l=1}^{L}\alpha_l(t)A_l$ with $A_l$ being Pauli strings). From Corollary 9 of~\cite{berry2020time}, one can easily show that the subsequent gate complexity now changes to $\mathcal{O}\big(\frac{\norm{\alpha}_{1,1}^2}{\epsilon} g_p\big)$, where $g_p$ is the gate complexity for Pauli exponential implementation, following
\begin{equation}
\int_{0}^{T}\mathrm{d}\tau\sum_{l=1}^{L}\alpha_l(t)\norm{A_l}_{\infty}
    =\norm{\alpha}_{1,1}.
\end{equation}
The URCC framework starts with the Dyson series expansion for the time-ordered evolution operator
\begin{align}
U=\mathcal{T}\exp\left[-i\int_{0}^{T} dtA(t)\right]=\sum_{l=0}^\infty\mathcal{T}\int d\bm{t} (-i)^{l}\prod_{l'=1}^lA(t_{l'}).
\end{align}
Embedding the Dyson series into a Poisson-weighted estimator with rate $\lambda>0$ gives the exact identity
\begin{align*}
U = e^{\lambda} \sum_{l=0}^{\infty} \underbrace{\frac{\lambda^l e^{-\lambda}}{l!}}_{\text{Poisson weight}} \sum_{p} \underbrace{\frac{l!}{\lambda^l} \mathcal{T} \int dt \prod_{l'=1}^l A_{p_{l'}}(t_{l'})}_{\text{Pauli String dist}(l; p)} \times \underbrace{(-i)^l \hat{\sigma}_{p_1} \cdots \hat{\sigma}_{p_l}}_{\text{Pauli String}(p)}
\end{align*}
This expansion has a profound interpretation: The evolution is equivalent to averaging over \textit{Pauli paths} weighted by their time-ordered interaction strengths. Each path corresponds to applying Pauli operators $\hat{\sigma}_{p_k}$ at randomly sampled times $t_k$, forming a continuous-time quantum walk. Nevertheless, direct sampling from the discrete distribution is computationally demanding. URCC solves this through a continuous embedding:
\begin{align*}
&\text{Pauli String dis}(l; p) \times \text{Pauli String}(p) = \frac{l!}{\lambda^l} \mathcal{T} \int d\bm{t} \left[ \prod_{l'=1}^l A_{\text{tot}}(t_{l'}) \right] \left[ \prod_{l'=1}^l \underbrace{\sum_{p} \frac{A_p(t_{l'})}{A_{\text{tot}}(t_{l'})} (-i \hat{\sigma}_p)}_{Q(t_{l'})} \right]
\end{align*}
The sampling algorithm transforms quantum dynamics into a classical stochastic process: (1) Generate interaction times $t_k$ via Poisson process with rate $A_{\text{tot}}(t)$; (2) At each $t_k$, select $\hat{\sigma}_p$ with probability $\propto \norm{A_p(t_k)}$; (3) Apply operators in chronological order (later times first).

However, the raw Dyson series has prohibitive variance $\mathcal{O}(e^{2\norm{\boldsymbol{\alpha}}_{1,1}})$. URCC achieves polynomial variance through algebraic restructuring to make the total normalization factor satisfy $C = 1+O(\norm{\boldsymbol{\alpha}}^2_{1,1}/r)$. Consequently, the sampling error is
\begin{align}
\epsilon = \mathcal{O}\left(\frac{C^2}{\sqrt{S}}\right)=\mathcal{O}\left(\frac{1}{\sqrt{S}}\Big(1+\mathcal{O}(\norm{\boldsymbol{\alpha}}^2_{1,1}/r^2)\Big)^{2r}\right).
\end{align}
To achieve accuracy $\epsilon$, it suffices to set $S=\mathcal{O}(1/\epsilon^2)$ and $r=\mathcal{O}(\norm{\boldsymbol{\alpha}}^2_{1,1})$.

\begin{theorem}[Observable-Driven Overhead]\label{unbiased_overhead}
Replacing the c-qDrift subroutine in Theorem~\ref{thm:ancilla_free_obs} with the URCC framework, observable $\mathcal{O}$ can be estimated to precision $\epsilon$ with probability at least $1-\delta$, by choosing the number of samples $S=\mathcal{O}(1/\epsilon^2)$ and
$r=\mathcal{O}\left(\norm{\boldsymbol{\alpha}}^2_{1,1} q^2 g_p\right)$ 
segments in each circuit without the dependence on $\epsilon$.
\end{theorem}

\section{Symmetric random-LCHS}
\label{sec:symmetry_sampling}
In this section, we develop a symmetry-aware importance sampling scheme based on physDrift for non-Hermitian systems that preserves key physical symmetries during the random-LCHS evolution. The approach consists of two key components: (1) identification of conserved quantities in the Hamiltonian models, and (2) construction of a symmetry-respecting sampling distribution for the randomized protocol.

\subsection*{Part 1}
The ability to solve the time evolution of a general matrix $A(t)$ enables the study of a much wider class of Hamiltonians through the Schr\"odinger equation, including those that are not Hermitian. Non-Hermitian Hamiltonians have recently attracted significant attention in quantum science, as they capture novel behaviors that do not arise in the Hermitian setting~\cite{ashida2020non,minganti2020hybrid,naghiloo2019quantum}. Our random-LCHS method efficiently generalizes to large non-Hermitian Hamiltonians, making it promising for near-term and early fault-tolerant devices. For the rest of the sections, we will focus on the Pauli models, for which the error bound is derived by directly replacing the norm from Theorem~\ref{thm:td_homo_random} to Eq.~\ref{pauli},

\begin{theorem}[Random-LCHS for Pauli-based Homogeneous ODEs]\label{thm:tid_homo_random}
    Consider the homogeneous ODE with $b(t) \equiv 0$ and $A \in \mathbb{C}^{2^{n}\times 2^{n}}$ that is time-independent, which admits the decomposition into Pauli strings.
    Then, there exists a quantum algorithm that prepares an $\epsilon$-approximation of the state $\ket{u(T)}$, with $\mathcal{O}\left( q \right)$
        queries to the state preparation oracles and gate complexity
    \begin{equation}
        \widetilde{\mathcal{O}} \left( \frac{ q_0^2g_p\norm{\boldsymbol{\alpha}}_{1}^2T^2}{\epsilon} \right),
    \end{equation}
    where $q_0=\frac{\|u_0\|_1}{\|u(T)\|_1}$ and $g_p$ is the gate required for a Pauli exponential.
\end{theorem}

Specifically, we consider two Hamiltonian models: the simple transverse 2D Ising model with a dissipative term and the Hatano-Nelson model. Both are improved numerically with the physical conserving property based on important sampling---symmetric random-LCHS. 

The physical intuition behind this technique is based on the observation that 
\begin{equation}
\label{eq:eta1}
i\frac{d}{dt}\langle\psi(t)|\eta|\psi(t)\rangle=\langle\psi(t)|\eta A-A^\dagger\eta|\psi(t)\rangle=0,
\end{equation}
where the linear operator $\eta$ is a constant of motion if and only if $\langle\psi(t)|\eta|\psi(t)\rangle=\mathrm{Tr}[\eta\rho_\psi(t)]$ remains constant for any arbitrary state $|\psi\rangle$ (or a density matrix $\rho_\psi$). Due to the linearity of the constraint in Eq.~\ref{eq:eta1}, without loss of generality, we can choose $\eta$ to be a Hermitian matrix. When $A=A^\dagger$, the observable conservation is therefore equivalent to commutation, just as expected. For a non-Hermitian Hamiltonian $A \neq A^\dagger$, the conserved Hermitian operators $\eta_k$ satisfy the intertwining relation:
\begin{equation}
\label{eq:general_intertwine}
\eta_k A = e^{i\phi} A^\dagger \eta_k
\end{equation}

We can explicitly construct these operators using a recursive procedure~\cite{ruzicka2021conserved}:
\begin{theorem}[Recursive Symmetry Construction]\label{recur}
For an $n$-dimensional Hamiltonian $A$ that is invariant under combined operations of the general parity and time-reversal ($\mathcal{PT}$) symmetry, the conserved operators $\{\eta_k\}_{k=1}^n$ satisfy:
\begin{align}
\eta_1 &= \mathcal{P} \\
\eta_{k+1} &= e^{i\phi/2} \eta_k A, \quad 1 \leq k \leq n-1
\end{align}
where $\mathcal{P}$ is the linear part of the $\mathcal{PT}$ operator.
\end{theorem}

\begin{proof}
Start with $\eta_1 = \mathcal{P}$ which satisfies $\mathcal{P}A = A^\dagger\mathcal{P}$ for transpose-symmetric Hamiltonians. The recursive relation follows by induction:
\begin{align*}
\eta_{k+1}A &= (e^{i\phi/2} \eta_k A) A \\
&= e^{i\phi/2} \eta_k A^2 \\
&= e^{i\phi/2} (e^{i\phi} A^\dagger \eta_k) A \quad \text{(by induction hypothesis)} \\
&= e^{i\phi(1 + 1/2)} A^\dagger (\eta_k A) \\
&= e^{i\phi_{}} A^\dagger \eta_{k+1}
\end{align*}
The phase $\phi_k$ is determined by the specific antilinear $\mathcal{PT}$ symmetry. For standard $\mathcal{PT}$ symmetry $\phi = 0$, for anti-$\mathcal{PT}$ symmetry $\phi = \pi$, and for anyonic symmetries $\phi \in (0,2\pi)$.
\end{proof}

\subsection*{Part 2}
After presenting the theoretical analysis, we validated the symmetry-protected sampler with numerical experiments on two representative models: the non-Hermitian Transverse-Field Ising Model (TFIM) and the Hatano-Nelson (HN) model. For each model, we compared the standard random-LCHS baseline against the symmetry-protected random-LCHS implementation. The protected implementation retains the same sampling marginals as the baseline but, for each sampled block, reorders the sampled multiset so as to alternate terms that tend to increase and decrease the monitored conserved observable. For every value of the sampling budget $r$, we ran multiple independent trials (with different random number generator seeds), computed the state error, and reported the mean and standard deviation across trials to visualize the average behavior and sampling variance as $r$ increases.

\vspace{0.2cm}
\noindent\textbf{Complex TFIM Model:} The Transverse-Field Ising Model with an imaginary longitudinal field has become a paradigmatic many-body example in PT-symmetric quantum mechanics~\cite{bender1998real,ashida2020non}, illustrating how simple lattice models can exhibit non-analytic symmetry-breaking transitions. Consider the explicit model:
\begin{equation}\label{eq:TFIM_model}
    A = H_{\text{TFIM}} + i\gamma\sum^n_{i=1}Z_i  = -J\sum_{i=1}^{n}Z_{i}Z_{i+1} - g \sum_{i=1}^{n}X_{i} + i\gamma\sum^n_{i=1}Z_i,
\end{equation}
where $J$ is the nearest-neighbor interaction strength, and $g$ is the coupling coefficient. It is evident that the norm of the state decays depending on the instantaneous magnetization along the $Z$-axis. This bias drives the system preferentially toward configurations with predominantly $Z_i=+1$ if $\gamma>0$, or $Z_i=-1$ if $\gamma<0$, which can be interpreted as a coherent coupling of each spin to an effective reservoir. Unlike a Lindblad formalism, where dissipation is encoded in jump operators and leads to mixed states, all amplification and decay remain within a pure-state, non-Hermitian Hamiltonian framework here. As $\gamma$ increases from zero, the spectrum evolves from entirely real to partially complex at a critical $\gamma_c$. Beyond $\gamma_c$, many-body modes exhibit strong non-unitary dynamics.

We decompose $A = L + iH$ with:
\begin{align*}
L = -J\sum_{i=1}^{n-1} Z_i Z_{i+1} - g \sum_{i=1}^n X_i, \quad
H = \gamma \sum_{i=1}^n Z_i
\end{align*}
The spectral norm $\|A\|$ is bounded by:
\begin{align*}
\|A\| \leq \|H\| + \|L\| \leq \left\|\gamma\sum_{i=1}^n Z_i\right\|+\left\|J\sum_{i=1}^{n-1} Z_i Z_{i+1}\right\| + \left\|g\sum_{i=1}^n X_i\right\| \leq \gamma n+J(n-1) + g n \approx \mathcal{O}\big(n\big).
\end{align*}
Choose the linear parity operator to be global spin flip $\mathcal P = \prod_{i=1}^n X_i$, and time reversal to be complex conjugation $\mathcal T = K$ (the usual antilinear operation sending $i\mapsto -i$ in the computational basis). We verify the invariance of each term in Eq.~\ref{eq:TFIM_model} under the combined action $\mathcal{PT}$.

The adjoint action of $\mathcal{P}$ on single‑site Pauli operators is
\begin{align*}
\mathcal P\,X_i\,\mathcal P^{-1} = X_i,\qquad
\mathcal P\,Z_i\,\mathcal P^{-1} = -Z_i,\qquad
\mathcal P\,Y_i\,\mathcal P^{-1} = -Y_i.
\end{align*}
Complex conjugation $K$ leaves $X, Z$ invariant in the real computational representation and sends $i\mapsto -i$ (hence $Y\mapsto -Y$ and explicit imaginary prefactors change sign). Apply $\mathcal P$ and then $\mathcal T$ term by term:
\begin{itemize}
\item The Ising coupling: $\mathcal P\big( -J\sum Z_iZ_{i+1}\big)\mathcal P^{-1} = -J\sum Z_iZ_{i+1}$ (two sign flips cancel) and $K$ leaves it unchanged.
\item The transverse field: $\mathcal P\big(-g\sum X_i\big)\mathcal P^{-1} = -g\sum X_i$, and $K$ leaves it unchanged.
\item The imaginary longitudinal field: $\mathcal P\big(i\gamma\sum Z_i\big)\mathcal P^{-1} = -i\gamma\sum Z_i$, and then $K$ sends $-i\mapsto +i$, restoring $+i\gamma\sum Z_i$.
\end{itemize}
Combining these results yields
\[ (\mathcal{PT})\,A\,(\mathcal{PT})^{-1} = A, \]
so $A$ is $\mathcal{PT}$‑symmetric for real $J,g,\gamma$ with the chosen $\mathcal P,\mathcal T$. Equivalently, one may write the intertwining relation $\mathcal{P} A = A^\dagger \mathcal{P}$, which is the operator identity used to build conserved intertwining operators.

From Theorem~\ref{recur} one constructs a tower of intertwining operators $\eta_n := \mathcal P\,A^{\,n}, n=0,1,2,\dots$,
each obeying $\eta_n A = A^\dagger\eta_n$. However, we will focus only on the first-order symmetry here, leaving higher-order conservation for future work. The goal is to approximate $e^{-AT}$ by a randomized product formula that preserves the $\mathcal{PT}$ structure, thereby reducing effective error growth from anti‑Hermitian pieces. The following algorithm is a principled modification of qDrift that enforces $\eta$‑pairing at the sampled step level.

\begin{algorithm}[H]
\caption{Importance-sampled LCHS discretization with symmetry-protected qDrift}
\label{alg:outer-lchs-qdrift}
\KwIn{Pre-resampled outer samples $\{k_m\}_{m=1}^N$ from the importance-sampling kernel, local decompositions $\{\ell_p\}, \{h_q\}$, symmetric operator $\eta$, total time $T$, inner qDrift steps $M_{\mathrm{in}}$.}
\KwOut{Approximate propagator $U(T)$}

$U \leftarrow I$, \quad $\tau \leftarrow T/N$\;
\For{$m \leftarrow 1$ \KwTo $N$}{
  Form $B(k) := kL + H$\;
  \For{$r \leftarrow 1$ \KwTo $M_{\mathrm{in}}$}{
    Construct local term list $\{b_j\} = \{k\ell_p\} \cup \{h_q\}$\;
    \For{$j \leftarrow 1$ \KwTo $|\{b_j\}|$}{
      compute $\widetilde b_j \leftarrow \eta\, b_j \,\eta^{-1}$\;
      compute weight $w_j \leftarrow \|b_j\| + \|\widetilde b_j\|$\;
    }
    $\Lambda \leftarrow \sum_j w_j$\;
    Sample index $j$ with probability $p_j = w_j/\Lambda$\;
    Apply paired short evolution:
    \[ U \leftarrow \exp\!\Big(-i\frac{\tau}{2M_{\mathrm{in}}}\widetilde b_j\Big)\,
             \exp\!\Big(-i\frac{\tau}{2M_{\mathrm{in}}} b_j\Big)\,U. \]
  }
}
\Return $U$\;
\end{algorithm}

\begin{figure}
    \centering
    \includegraphics[width=\linewidth]{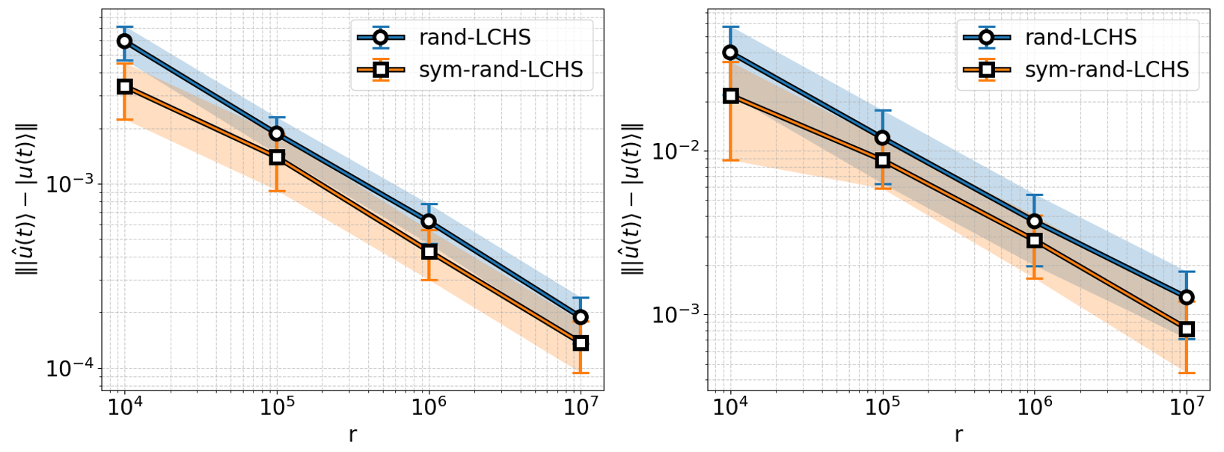}
    \caption{The log-log plots show mean error (with standard-deviation error bars) as a function of the sampling budget $r$. For both the TFIM (left) and the HN (right) benchmarking models, the mean errors decrease as $r$ increases, demonstrating the expected convergence of randomized-LCHS. The symmetry-protected sampler typically reduces the empirical error prefactor.
}
    \label{fgr:expt}
\end{figure}

\begin{figure}
    \centering
    \includegraphics[width=\linewidth]{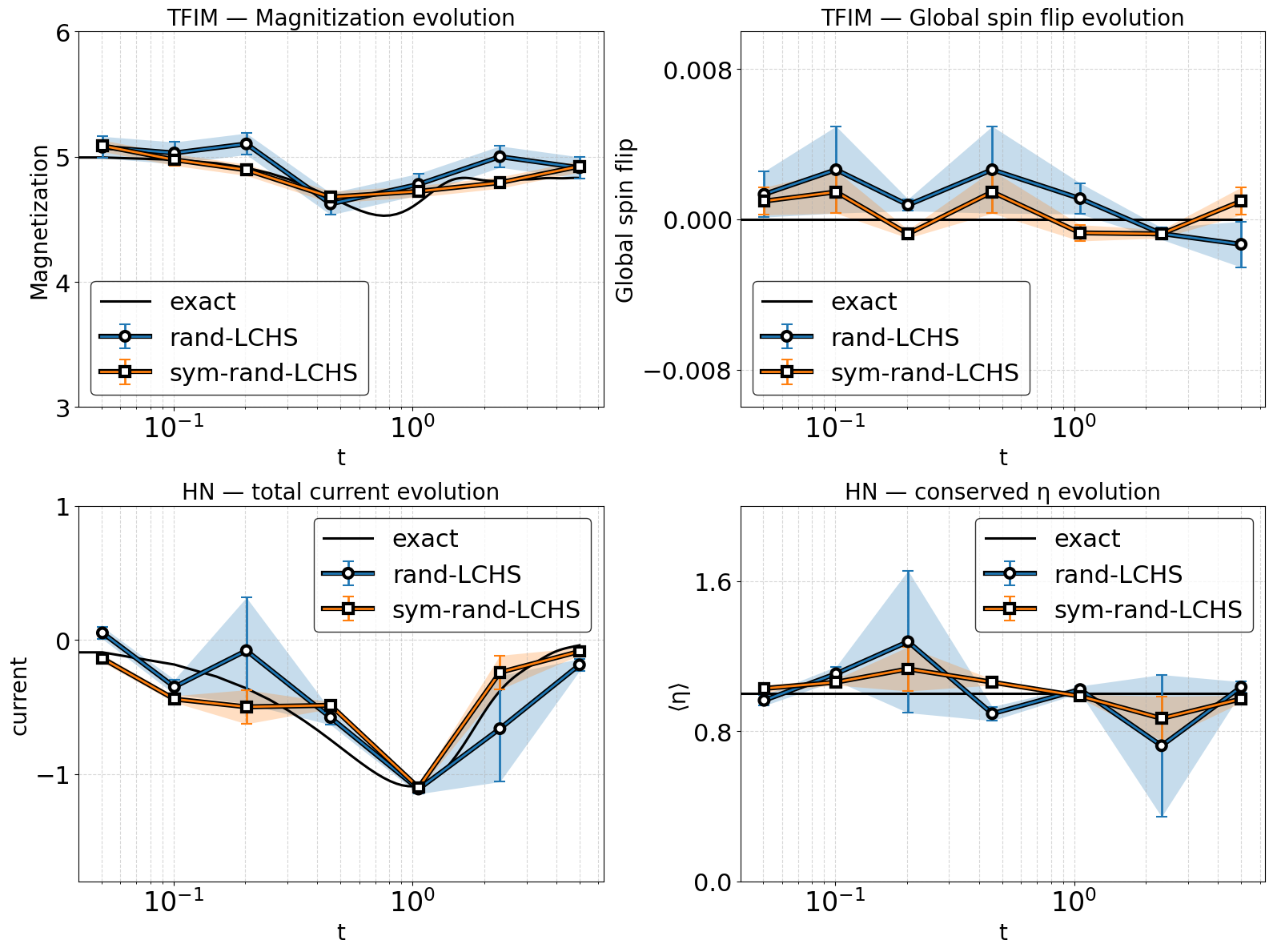}
    \caption{Benchmarking results for the TFIM (top row) and the HN (bottom row) models. The time span shown is five seconds (x-axis on a logarithmic scale). The top-left panel displays the magnitization evolution over the simulated time, and the top-right panel shows the global spin  $\prod_i \sigma_{x_i}$ evolution. The bottom-left panel plots the current in the HN model (briefly described in the \textbf{Benchmarking Results} section), while the bottom-right panel shows the conserved observable $\eta$ derived in the main text. Across all benchmarking experiments, the rand-LCHS results closely emulate the exact evolution trajectories, and the curves produced with symmetry constraints exhibit further improvement.
}
    \label{fgr:ob}
\end{figure}

\vspace{0.2cm}
\noindent\textbf{Interacting Hatano-Nelson Model:} A paradigmatic non‑Hermitian lattice model exhibiting non‑reciprocity and the non‑Hermitian skin effect is the Hatano–Nelson (HN) model~\cite{hatano1996localization,kunst2018biorthogonal}. We focus on the interacting fermionic variant on an open one‑dimensional chain of length $N$. The Hamiltonian is
\begin{equation}\label{eq:HN_fermionic}
H_{\mathrm{HN}} \,=\, \sum_{j=1}^{N-1} \Big[(J+\gamma)\,c^{\dagger}_{j+1}c_{j} + (J-\gamma)\,c^{\dagger}_{j}c_{j+1}\Big] \, + \, \sum_{i<j} V_{ij}\, n_i n_j,
\end{equation}
where $c_j,c_j^\dagger$ are fermionic annihilation/creation operators, $n_j=c_j^\dagger c_j$, and $J,\gamma\in\mathbb R$ set the symmetric and asymmetric hopping amplitudes respectively. The quartic interaction $V=\sum_{i<j}V_{ij}n_i n_j$ encodes density-density coupling between sites (e.g., nearest-neighbour or longer range).

For quantum simulation, it is convenient to map the fermions to spins using the Jordan-Wigner transform
$$ c_j = \Big(\prod_{m<j} Z_m\Big) \frac{X_j - iY_j}{2},\qquad n_j=\frac{1-Z_j}{2}, $$
which yields a Pauli‑string representation. The two‑site hopping piece of Eq.~\ref{eq:HN_fermionic} maps to
\begin{equation}\label{eq:HN_pauli}
H_{\mathrm{HN},j} \,=\, \frac{J}{2}(X_jX_{j+1}+Y_jY_{j+1}) \, - \, \frac{i\gamma}{2}(Y_jX_{j+1}-X_jY_{j+1}),
\end{equation}
so we again separate Hermitian and anti‑Hermitian components $H_{L,j}$ and $H_{A,j}$ when useful.

Following the TFIM model, we check if the HN model satisfies the $\mathcal{PT}$ symmetry. First, let us choose the parity operator to be spatial reflection about the centre of the chain,
$$ \mathcal P : j\mapsto N+1-j, \qquad \mathcal P\,c_j\,\mathcal P^{-1} = c_{N+1-j}. $$
Time reversal is again taken to be complex conjugation $\mathcal T=K$ (antilinear). Under the linear parity $\mathcal P$ the two hopping terms in Eq.~\ref{eq:HN_fermionic} are exchanged:
\begin{align*}
\mathcal P\,\big((J+\gamma)c^{\dagger}_{j+1}c_j\big)\,\mathcal P^{-1} &= (J+\gamma)c^{\dagger}_{N-j}c_{N+1-j},\\
\mathcal P\,\big((J-\gamma)c^{\dagger}_{j}c_{j+1}\big)\,\mathcal P^{-1} &= (J-\gamma)c^{\dagger}_{N+1-j}c_{N-j}.
\end{align*}
Since $\gamma$ is real, the subsequent complex conjugation $K$ does not change these real coefficients. Thus the combined $\mathcal{PT}$ action maps $H_{\mathrm{HN}}(\gamma)\mapsto H_{\mathrm{HN}}(-\gamma)$ (the asymmetric piece flips sign under parity while $K$ leaves real amplitudes unchanged). Therefore
\begin{equation*}
(\mathcal{PT})\,H_{\mathrm{HN}}(\gamma)\,(\mathcal{PT})^{-1} = H_{\mathrm{HN}}(-\gamma) \neq H_{\mathrm{HN}}(\gamma)
\end{equation*}
for generic $\gamma\neq 0$. Consequently, the interacting Hatano-Nelson model is \emph{not} $\mathcal{PT}$‑symmetric in the conventional sense (except trivially when $\gamma=0$).

However, we can still find a positive-definite intertwiner $\eta$ satisfying the pseudo-Hermiticity relation
\begin{equation}\label{eq:pseudo_def}
\eta\,H_{\mathrm{HN}} = H_{\mathrm{HN}}^{\dagger}\,\eta.
\end{equation}
Define the non-unitary site-dependent gauge
\begin{equation}\label{eq:S_def}
S = \exp\Big(\kappa\sum_{j=1}^N j n_j\Big),
\qquad n_j=c_j^\dagger c_j,
\end{equation}
with real parameter $\kappa$ to be chosen. The similarity action is
\begin{equation}\label{eq:S_action}
S\,c_j\,S^{-1} = e^{-j\kappa} c_j, \qquad S\,c_j^\dagger\,S^{-1} = e^{j\kappa} c_j^\dagger.
\end{equation}
Hence, a hopping operator transforms as
\begin{align*}
S\,c^{\dagger}_{j+1}c_j\,S^{-1} = e^{\kappa} c^{\dagger}_{j+1}c_j, \qquad
S\,c^{\dagger}_j c_{j+1}\,S^{-1} = e^{-\kappa} c^{\dagger}_j c_{j+1}.
\end{align*}
Applying $S$ to the kinetic part of Eq.~\ref{eq:HN_fermionic} gives
\begin{align*}
S\Big[ (J+\gamma)c^{\dagger}_{j+1}c_j + (J-\gamma)c^{\dagger}_j c_{j+1}\Big]S^{-1}
= (J+\gamma)e^{\kappa} c^{\dagger}_{j+1}c_j + (J-\gamma)e^{-\kappa} c^{\dagger}_j c_{j+1}.
\end{align*}
Choose $\kappa$ so that these transformed amplitudes are equal, i.e.
\begin{equation}\label{eq:kappa_condition}
(J+\gamma)e^{\kappa} = (J-\gamma)e^{-\kappa} \;\Longrightarrow\; e^{2\kappa} = \frac{J-\gamma}{J+\gamma}.
\end{equation}
This requires $|\gamma|<|J|$ so the right-hand side is positive; under this condition define the real hopping amplitude
\begin{equation}\label{eq:t_def}
\tilde J := (J+\gamma)e^{\kappa} = (J-\gamma)e^{-\kappa} = \sqrt{(J+\gamma)(J-\gamma)}.
\end{equation}
Consequently
\begin{equation}\label{eq:SHS_inv}
S\,H_{\mathrm{HN}}\,S^{-1} = \tilde J\sum_{j=1}^{N-1}\big(c^{\dagger}_{j+1}c_j + c^{\dagger}_j c_{j+1}\big) + V,
\end{equation}
which is manifestly Hermitian (the interaction $V$ commutes with $S$ because it depends only on the number operators $n_j$). Let $h:=S H_{\mathrm{HN}} S^{-1}$. Since $S H_{\mathrm{HN}} S^{-1}$ is Hermitian, we have
$$
h = h^\dagger=(S H_{\mathrm{HN}} S^{-1})^\dagger=(S^{-1})^\dagger H_{\mathrm{HN}}^\dagger S^\dagger.
$$
Equating the two expressions for $h$ and multiplying on the left by $S^\dagger$ and on the right by $S$ yields
$$
S^\dagger S\, H_{\mathrm{HN}}=H_{\mathrm{HN}}^\dagger\, S^\dagger S.
$$
Defining
\begin{equation}\label{eq:eta_from_S}
\eta = S^\dagger S =\exp\!\Big(2\kappa \sum_{j=1}^N j\, n_j\Big),
\end{equation}
we obtain the pseudo-Hermiticity (intertwining) relation
$$
\eta\, H_{\mathrm{HN}} = H_{\mathrm{HN}}^\dagger\, \eta.
$$
Since $S$ is invertible, $\eta=S^\dagger S$ is positive definite.

\vspace{0.2cm}
\noindent\textbf{Benchmarking Results:} We begin with the primary performance metric, the final-state error
$\|\hat{u}(t) - u(t)\|$, computed for each sampler as a function of the number of samples $r$ in~\fgr{expt}. Across both benchmark models, the symmetry-protected sampler (sym-rand-LCHS) exhibits a clear and consistent reduction in finite-sample error relative to the baseline, rand-LCHS protocol. Averaging over the tested samplings and the $40$ independent experiments with the same parameter settings (all plots show the sample mean $\pm$ the standard deviation), we observe the following empirical improvements in the final-state metric: in the TFIM model (system size $n=5$, final evolution time $T=2$, $J=1.0$, $g=0.5$, $\gamma=0.3$), the symmetry-protected sampler attains an average reduction in final-state error of approximately $31.7\%$, while in the HN chain (number of sites $L=16$, final evolution time $T=2$, $J=1.0$, $\gamma=0.3$, $V=0.5$) the corresponding average reduction is roughly $32.7\%$. On log-log axes of error versus the sample number $r$, both samplers follow the expected asymptotic decay with $r$; the effect of symmetry protection is to lower the finite-sample prefactor, producing smaller errors.

Before we move on to the observable experiments, we briefly introduce the definition and interpretation of the particle current used in the Hatano-Nelson benchmark. Starting from the Schr\"odinger equation for amplitudes $\psi_j$ in the Hermitian Hamiltonian after the transformation,
$
i\frac{d}{dt}\psi_j = \sum_k A_{j,k}\,\psi_k ,
$
the time derivative of the site occupation $\rho_j = |\psi_j|^2$ is
\begin{equation}
\frac{d}{dt}|\psi_j|^2
= \psi_j^* \dot{\psi}_j + \dot{\psi}_j^* \psi_j
= -i \sum_k \big( \psi_j^* A_{j,k} \psi_k - \psi_k^* A_{k,j} \psi_j \big).
\end{equation}
Restricting to nearest-neighbour hopping, the contribution between sites $j$ and $j+1$ becomes
$$
- i\big( \psi_j^* A_{j,j+1} \psi_{j+1} - \psi_{j+1}^* A_{j+1,j} \psi_j \big)
= -2\,\mathrm{Im}\!\big( \psi_j^* A_{j,j+1} \psi_{j+1} \big),
$$
which identifies the local current
$$
J_j = 2\mathrm{Im}\big(\psi_j^* A_{j,j+1}\, \psi_{j+1}\big).
$$
Because the original $A$ is non-Hermitian, total probability need not be conserved and a nonzero steady current may persist; the total current $J_{\mathrm{total}} = \sum_j J_j$ serves as a direct probe of non-reciprocal transport and as an order parameter for the non-Hermitian skin effect, where extensive current values indicate the formation of boundary-localized states. In our numerical experiments, we evaluate $\langle J_{\mathrm{total}}\rangle$ as a function of time during the evolution.

We now present the observable-level benchmarking as in~\fgr{ob} and relate it to the final-state improvements reported above. For each method, we compared the expectation values against the reference evolution and reported deviations in the conserved observables. In the TFIM case we measured the magnetization and its deviation. In the Hatano--Nelson case, we measured both the total particle current $\langle J_{\mathrm{total}}\rangle$. For both models, we performed a check for the constant evolution of the conserved quantities.

The symmetry-protected sampler reduces observable deviations in a manner consistent with the final-state metric. In the TFIM, the mean magnetization error suppressed is consistent with the reduction seen in the final-state error. In the Hatano-Nelson runs, the symmetry protection lowers both deviations at all tested values of $r$, yielding minor mean errors and, in many regimes, noticeably tighter error bars.

Qualitatively, the reduction in observable error is explained by the exact mechanism that improves the final state: symmetry-aware reordering of sampled terms reduces finite-sample bias in conserved quantities. It mitigates stochastic fluctuations that otherwise produce larger deviations in transport observables. In summary, with identical sampling budgets, the symmetry-protected rand-LCHS variant delivers a tangible finite-sample advantage, lowering final-state error by approximately $30\%$ on average and producing comparable reductions in physically relevant observable deviations, while preserving the expected asymptotic decay with $r$.

\section{Discussion and Outlook}\label{sec:discussion}
We have presented a systematical random-compilation framework for simulating time-dependent, inhomogeneous, linear non-unitary dynamics. The general random-LCHS algorithm achieves circuit-efficient simulation for the quantum-state evolution; the observable-driven random-LCHS estimates the observable with reduced sampling size and per-sample cost; and the symmetric random-LCHS employs the models' conserved quantities to reduce practical errors. Our work offers provable improvements on theoretical complexities and empirical enhancement on numerical benchmarks. We believe the work can contribute to developing the near-term and early fault-tolerant quantum algorithms as well as identifying end-to-end quantum scientific applications.

A promising direction for future work is to combine our random-LCHS strategy with the qFLO~\cite{watson2024randomly} protocol’s Richardson‐extrapolation step to mitigate the $1/\epsilon$ overhead arising from the c-qDrift layer to logarithmic complexity. Another subsequent advantage from qFLO is that we can easily reduce the $q^2$ scaling to linear dependence. This approach is similar to the unbiased sampling technique, which trades the circuit depth dependence on $\epsilon$ for more trials to perform Monte Carlo sampling. Concretely, one would apply the qFLO Richardson‐extrapolation procedure to the collection of expectation-value estimates obtained at each experiment step, thereby approximating the zero‐step‐size limit with only a logarithmic dependence on $\epsilon$.  

Finally, in addition to the models we considered here, an important physical extension of our non-Hermitian LCHS framework is to investigate the \emph{non-Hermitian skin effect}~\cite{okuma2020topological}, wherein a substantial fraction of eigenmodes localize exponentially at open boundaries, dramatically altering bulk-boundary correspondence and transport properties. The skin effect has been systematically reviewed in recent literature, including its one-dimensional minimal models~\cite{hatano1996localization}, phase transition~\cite{kawabata2023entanglement}, and higher-dimensional generalizations~\cite{zhang2022universal}. Its topological origin has been traced to point‐gap winding of the complex spectrum and symmetry protection (e.g.\ $\mathbb Z_2$ skin modes under time-reversal). With our random-LCHS, we can develop explicit circuit constructions, symmetry-respecting sampling schemes, and conduct numerical studies of skin effect phenomena within our non-Hermitian solver, paving the way for future work.

\begin{acknowledgments}
JPL acknowledges support from Innovation Program for Quantum Science and Technology (Grant No.2024ZD0300502), start-up funding from Tsinghua University and Beijing Institute of Mathematical Sciences and Applications.
\end{acknowledgments}

\bibliography{refs.bib}

\onecolumngrid
\clearpage

\appendix
\begin{center}
    \textbf{Supplementary Materials}
\end{center}

\section{Numerical Discretization}\label{sec:discretization}

This section details the numerical discretization of the LCHS formula $\mathcal{T} e^{-\int_0^t A(t) \ud t} = \int g(k) U(T,k) \ud k$ and its corresponding error bounds. We first address the homogeneous term and subsequently the inhomogeneous term.

\noindent\textbf{Homogeneous Term Discretization:} The integral is discretized via truncation and composite Gaussian quadrature. Define the kernel and the parameterized unitary:
\begin{align}\label{eqn:quadrature}
    g(k) &= \frac{1}{C_{\beta} (1-ik) e^{(1+ik)^{\beta}} }, \\
    U(T,k) &= \mathcal{T} e^{-i \int_0^T (kL(t)+H(t)) \ud t}.
\end{align}
The discretization procedure is:
\begin{equation}\label{eqn:LCHS_LCU_composite}
\begin{split}
    \mathcal{T} e^{-\int_0^T A(t) \ud t} &= \int_{\mathbb{R}} g(k) U(T,k) \ud k \approx \int_{-K}^{K} g(k) U(T,k) \ud k \quad \text{(Truncation)} \\
    &= \sum_{m = -K/h_1}^{K/h_1-1} \int_{mh_1}^{(m+1)h_1} g(k) U(T,k) \ud k \approx \sum_{m = -K/h_1}^{K/h_1-1} \sum_{q=0}^{Q-1} w_q g(k_{q,m}) U(T,k_{q,m}). \quad \text{(Quadrature)}
\end{split}
\end{equation}
Here, $h_1$ is the step size, $K/h_1$ is an integer, and $\{k_{q,m}, w_q\}$ are the nodes and weights for $Q$-point Gaussian quadrature on each subinterval.

\noindent\textbf{Truncation Error:} The error from restricting the integral to $[-K, K]$ is bounded by:
\begin{equation}
    \left\| \int_{|k|>K} g(k) U(T,k) \ud k \right\| \leq \frac{2^{\lceil 1/\beta \rceil+1} \lceil 1/\beta \rceil !}{C_{\beta} \cos(\beta\pi/2)^{\lceil 1/\beta \rceil} } \frac{1}{ K } e^{-\frac{1}{2}K^{\beta} \cos(\beta\pi/2) }.
\end{equation}
To ensure this error is below $\epsilon_{\text{trunc}}$, it suffices to choose:
\begin{equation} \label{eq:lambert_K}
    K = \Theta\left( \left(\log\left(\frac{B_\beta}{\epsilon_{\text{trunc}}}\right)\right)^{1/\beta} \right),
\end{equation}
where $B_\beta$ is a constant dependent on $\beta$. An optimized choice~\cite{pocrnic2025constant} uses the Lambert $W$ function: $$K = \Theta\left (\frac{2\beta}{\cos (\beta \pi /2)} W_0\left ( \left (\frac{B_\beta}{\epsilon_{\text{trunc}}} \right )^{1/\beta} \frac{\cos(\beta \pi /2)}{2\beta} \right ) \right )^{1/\beta}.$$

\noindent\textbf{Discretization Error:} The quadrature error is bounded by:
\begin{equation}
    \left\| \epsilon_{\text{disc}} \right\| \leq \frac{8 }{3C_{\beta}} K h_1^{2Q} \left(\frac{eT \norm{L(t)}_{\infty,\infty}}{2}\right)^{2Q}.
\end{equation}
To bound this error by $\epsilon_{\text{disc}}$, choose:
\begin{equation}
    h_1 = \frac{1}{eT  \norm{L(t)}_{\infty,\infty}}, \quad Q = \left\lceil \frac{1}{\log 4} \log\left( \frac{8K}{3C_{\beta}\epsilon_{\rm disc}} \right) \right\rceil.
\end{equation}
The total number of unitaries $M$ in the resulting LCU decomposition is:
\begin{equation}\label{M}
    M = \frac{2KQ}{h_1} = \mathcal{O}\left( T  \norm{L(t)}_{\infty,\infty} K Q \right) = \mathcal{O}\left( T  \norm{L(t)}_{\infty,\infty}\left(\log\left(\frac{1}{\epsilon_{\rm v}}\right)\right)^{1+1/\beta} \right),
\end{equation}
where $\epsilon_\text{trunc} = \epsilon_\text{disc} = \epsilon_\text{v}$.

The 1-norm of the LCU coefficients $\sum_{q,m} |c_{q,m}|$ is $\mathcal{O}(1)$, as it approximates the finite integral $\int_{-K}^{K} |g(k)| \ud k$. The final discretized evolution operator is expressed as:
\begin{equation}\label{eqn:LCHS_LCU}
    \mathcal{T} e^{-\int_0^T A(t) \ud t} \approx \sum_{j=0}^{M-1} c_j U(T,k_j).
\end{equation}

\noindent\textbf{Inhomogeneous Term Discretization:} The inhomogeneous term is discretized by a direct application of the LCHS formula, followed by the same composite Gaussian quadrature rules used for the homogeneous term. First, the formula is applied to the $k$-variable:
\begin{equation}
\int_0^T \mathcal{T}e^{-\int_s^T A(t')\ud t'} b(t) \ud t = \int_0^T \int_{\mathbb{R}} \frac{f(k)}{ 1-ik} U(T,t,k) b(t) \ud k \ud t \approx \int_0^T \sum_{j} c_j U(T,t,k_j) b(t) \ud t.
\end{equation}
A second quadrature is then applied to the time variable $s$, yielding the final discrete form:
\begin{equation}\label{eqn:discretization_inhomo_2}
\int_0^T \mathcal{T}e^{-\int_t^T A(t')\ud t'} b(t) \ud t \approx \sum_{j'=0}^{M'-1} \sum_{j=0}^{M-1} c'_{j'} c_{j} U(T,t_{j'},k_{j}) \ket{b(t_{j'})},
\end{equation}
where $c'_{j'}$ are the coefficients from the time quadrature and $\ket{b(t)}$ is the normalized vector.

The analysis of the truncation and discretization errors for this term closely mirrors that of the homogeneous case. The parameter $K$ scales as $\mathcal{O}\left( \left(\log\left(1+\frac{\|b\|_{1}}{\epsilon}\right)\right)^{1/\beta} \right)$, and the 1-norm of the time coefficients satisfies $\sum |c'_{j'}| = \mathcal{O}(\|b\|_{1})$. The total number of terms $M'$ in the double sum scales as:
\begin{equation}
M' = \widetilde{\mathcal{O}}\left( T (\Lambda+\Xi) \left(\log\left(1+\frac{\|b\|_{1}}{\epsilon}\right)\right)^{1/\beta} \right),
\end{equation}
where $\Lambda$ and $\Xi$ are the bounds on the higher-order time derivatives of $A(t)$ and $b(t)$, respectively. For a detailed derivation of the error bounds and parameter scaling, we refer the reader to the original source~\cite{an2023quantum}.

\section{Oracles and Implementation via Truncated Series}\label{sec:oraclestruncate}

This section outlines the oracles required to implement the LCU formula using a truncated Dyson series method and analyzes the resulting algorithm's complexity.

We assume access to the following oracles:
\begin{itemize}
    \item \textbf{Matrix Oracles (HAM-T):} Unitaries that block-encode the matrices $A(t)$, $L(t)$, and $H(t)$ at discretized time points. For example, for $A(t)$ and $\alpha_A \geq \norm{A(t)}_{\infty,\infty}$:
    \begin{equation}\label{eqn:oracle_A}
    (\bra{0}_a \otimes I) \text{HAM-T}_{A,q} (\ket{0}_a \otimes I) = \sum_{l=0}^{M_{\text{D}}-1} \ket{l}\bra{l} \otimes \frac{A(t_l)}{\alpha_A}.
    \end{equation}
    Oracles for $L(t)$ and $H(t)$ (with factors $\alpha_{A_L}, \alpha_{A_H}$) can be constructed from $\text{HAM-T}_{A,q}$ or provided directly.
    \item \textbf{Coefficient Oracles:} State preparation unitaries $O_{c,l}, O_{c,r}$ that prepare the LCU coefficients $c_j$ from the $k$-quadrature, and $O_{c',l}, O_{c',r}$ for the time quadrature coefficients $c'_{j'}$.
    \item \textbf{Node Oracle:} A unitary $O_k: \ket{j}\ket{0} \rightarrow \ket{j}\ket{k_j}$ that outputs the quadrature node $k_j$ for index $j$.
    \item \textbf{State Preparation Oracles:} $O_u: \ket{0} \rightarrow \ket{u_0}$ for the initial state and $O_b: \ket{j'}\ket{0} \rightarrow \ket{j'} \ket{b(s_{j'})}$ for the inhomogeneous term.
\end{itemize}
The core of the algorithm is to construct a SELECT oracle for the unitaries $U(T,k_j)$. The first step is to build a block-encoding of the Hamiltonian $k_j L(t) + H(t)$ for a given $j$. Starting from the state $\ket{j}\ket{l}\ket{0}_k\ket{0}_R\ket{0}_a\ket{\psi}$:
(1) Apply $O_k$ to get $\ket{k_j}$; (2) Apply a controlled rotation to create a superposition selecting between the $L$ and $H$ terms based on their weights; (3) Apply controlled $\text{HAM-T}_{L,q}$ and $\text{HAM-T}_{H,q}$ oracles; (4) Uncompute the rotation and apply a second rotation to normalize the block-encoding factor to the worst-case value $\alpha_{A_L} K + \alpha_{A_H}$.
This procedure implements an oracle $\text{HAM-T}_{kL+H,q}$ satisfying:
\begin{equation}\label{eqn:algorithm_hamt_kLH}
(\bra{0}_{a'}\otimes I) \text{HAM-T}_{kL+H,q} (\ket{0}_{a'} \otimes I) = \sum_{j,l} \ket{j}\bra{j} \otimes \ket{l}\bra{l} \otimes \frac{(k_j L+H)(t_l)}{\alpha_{A_L} K + \alpha_{A_H}}.
\end{equation}
Using the above HAM-T oracle, the truncated Dyson series method implements a block-encoding $W_j \approx U(T,k_j)$. The SELECT oracle is then $\text{SEL} = \sum_{j} \ket{j}\bra{j} \otimes W_j$. The homogeneous time evolution operator is block-encoded by:
\begin{equation}
(O_{c,l}^{\dagger} \otimes I) \text{SEL} (O_{c,r}\otimes I).
\end{equation}
Applying this to $\ket{0}O_u\ket{0}$ yields a state proportional to the homogeneous solution plus garbage.

The unitary $U(T,s_{j'},k_j)$ is implemented similarly, using oracles for the time-shifted Hamiltonians $\mathbb{I}_{s'\geq s} L(t')$ and $\mathbb{I}_{s'\geq s} H(t')$. The full inhomogeneous term is prepared by:
\begin{equation}
(O_{c',l}^{\dagger} \otimes O_{c,l}^{\dagger} \otimes I) \widetilde{\text{SEL}}' (O_{c',r} \otimes O_{c,r} \otimes I),
\end{equation}
where $\widetilde{\text{SEL}}'$ applies $W_{j,j'} \approx U(T,s_{j'},k_j)$ and then $O_b$. This yields a state proportional to the inhomogeneous solution.

To evaluate the matrix query complexity and the gate complexity, we need the following lemma from~\cite{low2018hamiltonian}.
\begin{lemma}[Hamiltonian simulation by a truncated Dyson series]\label{Thm:Compressed_TDS_Uniform}
	Let $H(t) : [0,t] \rightarrow \mathbb{C}^{2^{n_s}\times 2^{n_s}}$, and let it be promised that $ \norm{H(t)}_{\infty,\infty}\le\alpha$, and $\langle\|\dot H\|_\infty\rangle=\frac{1}{t}\int^{t}_{0} \left\|\frac{\mathrm{d} H(t)}{ \mathrm{d} s}\right\|_\infty \mathrm{d}s$. Let $\tau=t/\left\lceil 2\alpha t\right\rceil$ and assume $H_j(t)=H((j-1)\tau + s): s\in[0,\tau]$ is accessed by an oracle $\operatorname{HAM-T}_j$ with $M\in{\mathcal{O}}\left( \frac{t}{\alpha\epsilon}\left({\langle \|\dot{H}\|_\infty \rangle}  +{ \norm{H(t)}_{\infty,\infty}^2}\right)\right)$.  For all $|t|\ge 0$ and $\epsilon> 0$, an operation $W$ can be implemented with failure probability at most $\mathcal{O}(\epsilon)$ such that $\left\|W-\mathcal{T}\left[ e^{-i\int_0^t H(t) \mathrm{d} s}\right]\right\| \le \epsilon$ with the following cost.
	\begin{enumerate}
		\item Queries to all $\operatorname{HAM-T}_j$: $\mathcal{O}\left(\alpha t\frac{\log{(\alpha t/\epsilon)}}{\log\log{(\alpha t /\epsilon)}}\right)$,
		\item Qubits: $n_s + \mathcal{O}\left(n_a+\log{\left( \frac{t}{\alpha\epsilon}\left({\langle \|\dot{H}\| \rangle_\infty} +{\alpha^2}\right)\right)}\right)$,
		\item Primitive gates: \\$\mathcal{O}\left(\left(n_a +\log{\left( \frac{t}{\alpha\epsilon}({\langle \|\dot{H}\| \rangle_\infty}  +{\alpha^2})\right)}\right)\frac{\alpha t\log{(\alpha t/\epsilon)}}{\log\log{(\alpha t/\epsilon)}}\right)$.
	\end{enumerate}
\end{lemma}

We now present a complexity analysis for a general inhomogeneous ODE. 

\begin{theorem}\label{thm:complexity_inhomo} 
    Suppose that $L(t)$ is positive semi-definite on $[0, T]$, and we are given the oracles described in Section.\ref{sec:oraclestruncate}. 
    Let $ \norm{A(t)}_{\infty,\infty} \leq \alpha_A$ and define $\Lambda = \sup_{p \geq 0, t \in [0,T]} \|A^{(p)}\|_\infty^{1/(p+1)} $ and $\Xi = \sup_{ p\geq 0, t \in [0,T] } \|b^{(p)}\|_\infty^{1/(p+1)} $, where the superscript $(p)$ indicates the $p$th-order time derivative. 
    Then we can prepare an $\epsilon$-approximation of the normalized solution $\ket{u(T)}$ with $\Omega(1)$ probability and a flag indicating success, by choosing 
    \begin{equation}
         M = \mathcal{O}\left( \alpha_A T\left(\log\left(\frac{\|u_0\|_1+\|b\|_1}{\|u(T)\|_1 \epsilon}\right)\right)^{1+1/\beta} \right), \qquad M' = \widetilde{\mathcal{O}}\left( T (\Lambda+\Xi) \left(\log\left(\frac{1+\|b\|_1}{\|u(T)\|_1\epsilon}\right)\right)^{1+1/\beta} \right), 
    \end{equation}
    using
        \begin{equation}
            \widetilde{\mathcal{O}}\left( \frac{\|u_0\|_1+\|b\|_1}{\|u(T)\|_1} \alpha_A T \left(\log\left(\frac{\|u_0\|_1+\|b\|_1}{\|u(T)\|_1\epsilon}\right)\right)^{2+1/\beta}  \right)
        \end{equation}
        primitive gates with the $\text{HAM-T}$ oracle, $\mathcal{O}\left( \frac{\|u_0\|+\|b\|_{1}}{\|u(T)\|} \right)$ queries to the state preparation oracles $O_u$ and $O_{b}$ and, $            \widetilde{\mathcal{O}}\left(\log{\left( T^2\alpha_A(\Lambda+\Xi)\right)}\right)$ extra ancilla qubits.
\end{theorem}
\begin{proof}
The rest of the proof mainly follows from \cite{an2023quantum} except for the fact that we do not consider the query complexity to the $\text{HAM-T}_{kL+H,q}$ oracle. Instead, we consider the gate complexity for the block-encoding $\|V_j - U(T,k_j)\| \leq \epsilon_1$ for any $\epsilon_1 > 0$, which implements $\text{SEL} = \sum_{j=0}^{M-1} \ket{j}\bra{j} \otimes W_j$. Using Lemma.\ref{Thm:Compressed_TDS_Uniform},
    \begin{equation}
    \begin{split}
    &\mathcal{O}\left(\left(\log(M) +\log{\left( \frac{T({\langle \|\dot{A}\| \rangle}  +{(\alpha_{A_L}K+\alpha_{A_H})^2})}{(\alpha_{A_L}K+\alpha_{A_H})\epsilon_1}\right)}\right)\frac{(\alpha_{A_L}K+\alpha_{A_H}) T\log{((\alpha_{A_L}K+\alpha_{A_H}) T/\epsilon_1)}}{\log\log{((\alpha_{A_L}K+\alpha_{A_H}) T/\epsilon_1)}}\right)\\
    &= \widetilde{\mathcal{O}} \left( \alpha_A K T \log^2(T{\langle \|\dot{A}\| \rangle} \alpha_A/\epsilon_1) \right),
    \end{split}
    \end{equation}
We now have a quadratic relation in the logarithmic dependence. Then, $(O_{c,l}^{\dagger} \otimes I) \text{SEL} (O_{c,r}\otimes I)$ is a block-encoding of $\frac{1}{\|c\|_1}\sum_{j=0}^{M-1} c_j V_j$, and thus 
    \begin{equation}\label{eqn:complexity_proof_eq1}
        (O_{c,l}^{\dagger} \otimes I) \text{SEL} (O_{c,r}\otimes I) (O_u\otimes I)\ket{0}\ket{0} = \frac{1}{\|c\|_1} \ket{0} \left(\sum_{j=0}^{M-1} c_j V_j\right)\ket{u_0} + \ket{\perp}. 
    \end{equation}
    This step only needs $1$ query to $O_{c,l},O_{c,r},O_u$, and $\text{SEL}$. By the same argument, we can prepare a single copy of Eq.~\ref{eqn:algorithm_general_inhomo_part} using $\mathcal{O}(1)$ queries to $O_{c,l}, O_{c,r}, O_{c',l}, O_{c',r}, O_b$ and $\widetilde{\mathcal{O}} \left( \alpha_A K T \log^2(1/\epsilon_2) \right)$ queries to $\text{HAM-T}_L$ and $\text{HAM-T}_H$.
    Here $\epsilon_2$ is an upper bound on $\|V_{j,j'} - U(T,s_{j'},k_j)\|$. 
    Therefore, we can prepare a single copy of the solution by querying $O_{c,l}, O_{c,r}, O_{c',l}, O_{c',r}, O_u, O_b$ $\mathcal{O}(1)$ times with gate count
    \begin{equation}
        \widetilde{\mathcal{O}}\left( \alpha_A K T \log^2\left(\frac{1}{\min \left\{\epsilon_1,\epsilon_2\right\}}\right) \right).
    \end{equation}

   The determination of the choice of $\epsilon_1$ and $\epsilon_2$ follows the approach in~\cite{an2023linear} and our main text. The same choice applies to $ M$, $ M'$, and $K$. A single run of the algorithm queries $\text{HAM-T}_L$ and $\text{HAM-T}_H$ with gate complexity
    \begin{align}
        \widetilde{\mathcal{O}}\left( \alpha_A K T \left(\log\left(\frac{1}{\min \left\{\epsilon_1,\epsilon_2\right\}}\right)\right)^2 \right) &= \widetilde{\mathcal{O}}\left( \alpha_A K T \left(\log\left(\frac{ \|c\|_1(\|u_0\|_1 + \|c'\|_1)}{ \|u(T)\|_1 \epsilon}\right)\right)^2 \right) \\
        & = \widetilde{\mathcal{O}}\left( \alpha_A T \left(\log\left(\frac{ \|u_0\|_1 + \|b\|_1}{ \|u(T)\|_1 \epsilon}\right)\right)^{2+1/\beta} \right).
    \end{align}
    The repetition prefactor for the amplitude amplification and the ancilla analysis naturally follows from the main text as well. 
\end{proof}

\section{The Continuous qDrift Protocol}\label{review_cqdrift}
The continuous qDrift (c-qDrift) protocol is a randomized method for Hamiltonian simulation. For a time-dependent Hamiltonian $H(t)=\sum_{l=1}^{L}H_l(t)$, it constructs a quantum channel by sampling evolution times according to the norm of the Hamiltonian components:
\begin{equation}
\mathcal{U}_t(\rho):=\sum_{l=1}^{L}\int_0^T p_l(t) e^{-i H_l(t)/p_l(t)} \rho e^{i H_l(t)/p_l(t)} \ud t,
\end{equation}
where the probability distribution is given by:
\begin{equation}
p_l(t):=\frac{\norm{H_l(t)}_\infty}{\int_0^T \sum_l \norm{H_l(t)}_\infty \ud t}.
\end{equation}
A key result is the universality theorem, which demonstrates that this channel is equivalent to one defined by a single time-dependent Hamiltonian, $G(t)$. The diamond norm error between $\mathcal{U}_t$ and the true evolution channel $\mathcal{E}_t$ is bounded by:
\begin{equation}
\norm{\mathcal{E}_T-\mathcal{U}_T}_\diamond\leq 4\left(\int_0^T \sum_l \norm{H_l(t)}_\infty \ud t\right)^2.
\end{equation}
For error analysis, we note that the diamond norm provides a natural metric for comparing quantum channels. The Mixing Lemma establishes an important relationship between unitary and channel distances:
\begin{lemma}[Mixing lemma~\cite{berry2015hamiltonian}]
	\label{lem:diamond_bound}
	For any two unitary operators $U$ and $V$, with corresponding channels $\mathcal{U}(\rho)=U\rho U^\dagger$ and $\mathcal{V}(\rho)=V\rho V^\dagger$, we have:
	\begin{equation}
		\norm{\mathcal{U}-\mathcal{V}}_\diamond \leq 2\norm{U-V}_\infty.
	\end{equation}
\end{lemma}
This lemma enables a fair comparison between different simulation methods, whether they produce channels (such as c-qDrift) or unitaries (like the truncated Dyson series).

\end{document}